\documentclass[12pt]{article}

\usepackage{graphicx}
\usepackage{amsmath, amssymb}
\usepackage[numbers]{natbib}
\usepackage{authblk}

\usepackage{physics}

\usepackage{algorithm}
\usepackage{algorithmic}

\usepackage{newfloat}
\DeclareFloatingEnvironment[
    fileext=loc,
    listname={List of Circuits},
    name=Circuit,
    placement=tbhp,
]{circuit}
\usepackage{float}
\floatstyle{ruled}
\restylefloat{circuit}

\usepackage[colorlinks=true, linkcolor=blue, citecolor=blue, urlcolor=blue]{hyperref}

\usepackage{booktabs}
\usepackage{caption}
\usepackage{subcaption}
\usepackage{cleveref}
\usepackage{makecell}

\usepackage[letterpaper,margin=1in]{geometry}

\newcommand{\interval}{\sqcup}

\def\cN{\mathcal{N}}
\def\cR{\mathcal{R}}
\def\cE{\mathcal{E}}
\def\bR{\mathbb{R}}

\def\bZ{\mathbb{Z}}
\def\bF{\mathbb{F}}
\newcommand{\ted}{\textrm{Det}}
\usepackage{tikz-cd}
\usepackage{amsmath, amsthm,amssymb} 
\newtheorem{definition}{Definition}[section]

\title{A Topologically Fault-Tolerant Quantum Computer with Four Dimensional Geometric Codes
}
\author{David Aasen}
\author{Matthew Hastings}
\author{Vadym Kliuchnikov}
\author{Juan~M.~Bello-Rivas}
\author{Adam Paetznick}
\author{Rui Chao}
\author{Ben~W.~Reichardt}
\author{Matt Zanner}
\author{Marcus~P.~da~Silva}
\author{Zhenghan Wang}
\author{Krysta M. Svore}
\affil{Microsoft Quantum}

\date{\today}

\date{}

\begin{document}

\maketitle

\begin{abstract} 

Topological quantum codes are intrinsically fault-tolerant to local noise, and underlie the theory of topological phases of matter. 
We explore geometry to enhance the performance of topological quantum codes by rotating the four dimensional self-correcting quantum memory, and present codes targeted to both near-term and utility-scale quantum computers. 
We identify a full set of logical Clifford operations and with it design a universal fault-tolerant quantum architecture.   
Our design achieves single-shot error correction, significant reductions in required qubits, and low-depth logical operations. 
In turn, our proposed architecture relaxes the requirements for achieving fault tolerance and offers an efficient path for realization in several near-term quantum hardware implementations.
Our [[96,6,8]] 4D Hadamard lattice code has low weight-6 stabilizers and depth-8 syndrome extraction circuits, a high pseudo-threshold of $\sim 0.01$, and a logical error rate of $\sim 4 \times 10^{-7}$ per logical qubit per round of error correction at $10^{-3}$ physical error rate under a standard circuit-level noise model. 
A Clifford-complete logical gate set is presented, including a constructive and efficient method for Clifford gate synthesis.

\end{abstract}

\tableofcontents

\newpage

\section{Introduction}

Engineering a quantum computer capable of performing millions and more quantum operations requires quantum error correction and fault tolerance, as well as co-design of the hardware architecture with the implementation of the quantum error correcting code.
For quantum error correction and fault tolerance to enable deeper and more reliable computation, the quantum error correcting code must dramatically reduce the hardware failure rate, while also enabling efficient encoding of both information and operations.  
The code must not only enable a universal, fault-tolerant set of logical operations, but there must also be an efficient algorithm by which to compile into such operations.
Further, the code should possess a high rate of encoding, meaning few physical qubits are required per logical qubit, and desirably be single shot \cite{bombin2015single}, meaning only one round of syndrome extraction is sufficient for achieving the code distance in performance.

Topological quantum codes such as the two-dimensional toric codes are widely used in current hardware (see, e.g., \cite{acharya2024quantum,krinner2022realizing}), due to their planar layout and high threshold \cite{fowler2009high}. 
While such codes possess desirable logical operations by way of transversal logical operations or lattice surgery, and are well suited to hardware designs with a fixed grid connectivity, they exhibit a low encoding rate, requiring in turn $d^2$ physical qubits per logical qubit for code distance $d$ in the rotated case.
Further, they are not single-shot error correcting \cite{bombin2015single}, meaning they require a number of cycles for error correction that scales provably with the distance of the code.

Emerging hardware platforms with the ability to simulate all-to-all connectivity, such as neutral atoms, ion traps and photonics, among others, relax the need for codes to be geometrically local and dimension-constrained, and open an opportunity to consider a much broader class of codes with higher dimension and connectivity requirements (see, e.g., \cite{graham2023multiscale, Bluvstein_2023}).  
One such recently developed class of codes is the quantum low-density-parity-check (LDPC) codes, which exhibit high encoding rate.  It is an active area of research to make quantum LDPC codes efficient for quantum computing and accessible for near-term hardware implementation. Single-shot codes with known high thresholds and good performance that are capable of universal logical operations with efficient synthesis are highly prized for near-term hardware.  The 4D geometric codes presented herein stand out as a family of codes that combines the benefits of single-shot, high encoding rate, good performance, high threshold error rate, and efficient synthesis of universal logical operations.

We consider topological quantum codes, for which geometry has played a successful and important role for enhancing their performance theoretically \cite{freedman2002z2,hastings2021fiber}.  
Recently, Ref.~\cite{aasen2025topological} presented a general theory on enhancing topological codes by geometry. 
Here we focus on the $4$-dimensional torus and the change of its combinatorial geometry by rotating the standard lattices into general ones. 
Our $4$D geometric codes require only $6$-valent connectivity and possess high encoding rates, while being provably single shot error correcting.
We provide simulations of their performance for a range of code distances and show they exhibit a high threshold.
Importantly, we identify a universal set of fault-tolerant operations for the so-called Hadamard code and present an efficient synthesis algorithm for Clifford operations.
We show that fold-transversal gates paired with lattice surgery are Clifford complete and can be made into a universal set by state injection and distillation.
Our codes are well-suited for hardware implementations that can simulate all-to-all connectivity at low cost, such as neutral atoms, photonic platforms, and ion trap systems, and perform well at both near-term and scaled quantum hardware settings.

\section{Four dimensional geometric codes}
We begin with notation required to define the $4$D geometric codes.

\subsection{Notation}

\begin{enumerate}
    \item The Euclidean 4-space will be denoted as $\mathbb{R}^4$, whose 4-coordinates will be denoted by $\{x,y,z,w\}$ and ordered as $0,1,2,3$.
    \item An integer lattice will be denoted as $\Lambda$, whose matrix representatives will be denoted by $L$.  The rows of such matrices  $L$ will be vectors of $\mathbb{R}^4$.
    \item A $k$-cell for $k=0,1,2,3,4$ of the standard 4-cube $I^4,I=[0,1]$, will be denoted by $k$ empty holders of bits $\{0,1\}$ with the remaining $4-k$ coordinates with fixed bit values. The empty holder will be denoted by $\interval$ as in a Turing machine tape.
    \item A $k$-cell in the hypercubic lattice $\mathbb{Z}^4$ of the 4-cube sitting at a lattice point $p$ will be denoted as a 4-tuple consisting of $k$ empty bit holders $\interval$ and $4-k$ bit values with subscript $p$. The 4-cell sitting at the origin $(\interval \interval \interval \interval)_{(0,0,0,0)}$ is the standard unit cube $[0,1]\times [0,1]\times [0,1]\times [0,1]$.  A 2-cell or square is given by one of the following   
\begin{align}
&(0 0 \interval \interval)_p\\
&(0 \interval 0 \interval)_p\\
&(0 \interval \interval 0)_p\\
&(\interval 0 0 \interval)_p\\
&(\interval 0 \interval 0)_p\\
&(\interval \interval 0 0)_p
\end{align}
    \item A collection of $i$-cells will be denoted as $\{e^i_a\}$ for $i=0,1,2,3$ and indexed by $a$. Note that $0$-cells=vertices, $1$ cell=edges, $2$ cells=faces, $3$ cells=cubes, and $4$ cells=hypercubes. The dual cell of $e^i_a$ will be denoted as $\hat{e}^i_a$.

\end{enumerate}

Since our codes are topological, the formalism of $\mathbb{F}_2$-chain complexes for CSS codes will be used for our technical presentation \cite{bravyi2014homological}.

\subsection{Geometric enhancement of topological codes}\label{geometric}

CSS codes can be presented as a $3$-term $\mathbb{F}_2$-chain complex,

\begin{equation}
\begin{tikzcd}
C_2
\arrow[r, bend left=20, "\partial_2"] &
C_1 
\arrow[l, bend left=20, "\partial^T_2"]
\arrow[r, bend left=20, "\partial_1"]&
C_0
\arrow[l, bend left=20, "\partial^T_1"]
\end{tikzcd}
\end{equation}
and a CSS code will be called {\it topological} if the chain complex can be identified with the chain complex of the $\{2,1,0\}$ cells of a cellulation of some topological space, and furthermore that the logical code subspace are identified with the complex span of the first homology and the defining maps $H_Z$ and $H_X^T$ associated with $\partial_2$ and $\partial_1$.
The paradigmatic example is the toric code \cite{kitaev1997quantum}, where the topological space is the 2-torus $T^2$ with a cellulation to the square lattice with an $l\times l$ square unit cell.  More generally $\{2,1,0\}$ can be replaced by any three consecutive dimensions $\{i+1,i,i-1\}$. In the case of the 4D loop-only toric codes (whose elementary excitations have no particle-like ones, and the simplest are loops), $i=2$, so the chain complex consists of cubes, squares, and edges in $\bZ^4$ of the Euclidean 4-space $\bR^4$ that form part of a cellulation of a 4-torus $T^4$.  Conceptually, topological codes are simply ground states of topological phases of matter.
\footnote{
Two closely related approaches to fault-tolerance are topological quantum codes and topological quantum computing. On one hand, topological quantum computing utilizes the intrinsically fault-tolerant non-local degrees of freedom of topological phases, such as the ground state subspace of a collection of non-Abelian objects (see \cite{aasen2025roadmap} and references therein.) On the other hand, topological quantum codes actively realize fault-tolerance by measuring often commuting Pauli check operators, which is related to the local Hamiltonian terms of abelian topological phases. The latter approach does not require full control of non-abelian objects, instead relying on creating snap-shot topological states and performing operations fault-tolerantly.  A snap-shot topological state is a quantum state that coincides with a topological state of some topological phase of matter at a certain moment.  Then the evolution of the state completely diverges from the topological phase: no gapped Hamiltonians to protect the state, hence no intrinsic error resistance from the topological order.  The snap-shot topological states in our 4D geometric codes presented herein are from self-correcting quantum memories \cite{dennis2002topological}, and they may still retain some amount of error resistance from the topological order. A quantum computer can be regarded as a phase of matter, and by extension a fault-tolerant quantum computer would then be a synthetic topological phase.  This is an extension of the well-established characterization of traditional topological phases of matter: the ground states of a topological phase of matter are error correction codes \cite{kitaev2003fault,freedman2003topological}. In some sense, conceptually, this topologization of traditional fault-tolerance unifies the two topological approaches to fault-tolerant quantum computing.
}

\begin{figure}
    \centering
    \includegraphics[width=0.95\linewidth]{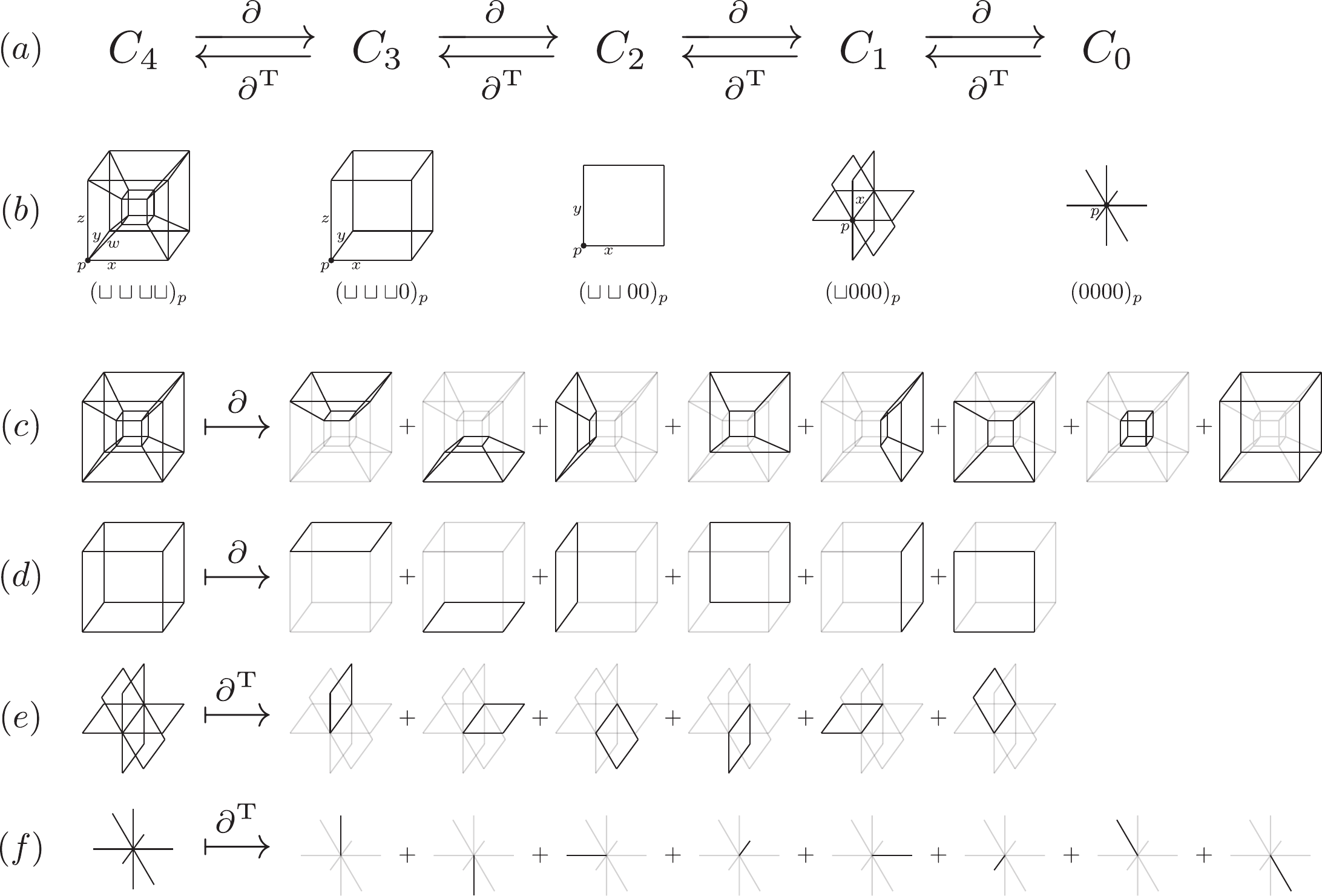}
    \caption{Summary of the 4D loop-only toric code. 
(a) The chain complex for the 4D hypercubic lattice, with vector spaces $C_k$ of $k$-cells ($k = 0,\ldots,4$), and boundary ($\partial$) and coboundary ($\partial^\top$) operators connecting them.
(b) Notation used to label cells, where coordinates indicate which directions are fixed ($0$) or span an interval ($\sqcup$) at a reference point $p$.
(c) The boundary of a 4-cell consists of eight 3-cells. The product of the corresponding 3-cell Z-stabilizers is the identity, indicating a redundancy.
(d) The boundary of a 3-cell consists of six 2-cells. The associated Z-stabilizer acts on the qubits living on these 2-cells.
(e) The coboundary of a 1-cell yields six 2-cells, defining the support of an X-stabilizer.
(f) The coboundary of a 0-cell gives eight 1-cells, producing a redundancy among the 1-cell X-stabilizers.
    }
    \label{fig:notation}
\end{figure}

We summarize the 4D loop-only toric code\footnote{also called (2,2)-toric code in the literature.} and the notation used in this manuscript in \Cref{fig:notation}.  The 4-cells (hypercubes) and 0-cells (vertices)  provide redundancies on the 3- and 1-cell (cube and edge) stabilizers, respectively, leading to the famed single-shot property. 

\begin{figure}
    \centering
    \includegraphics[width=0.3\linewidth]{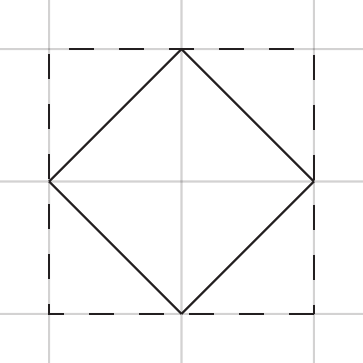}
    \caption{Geometry change by lattices: the distance of two diagonal vertices in a square is 2 in the Manhattan distance vs $\sqrt{2}$ in the Euclidean distance.  The $[[8,2,2]]$ toric code is shown in dashed lines, and the $[[4,2,2]]$ rotated toric code is shown in solid lines.  Both [[8,2,2]] and [[4,2,2]] are distance=2 codes, whose distances are the lengths of the shortest essential 1-cycles of the tori and their dual versions, respectively.}
    \label{fig:small_rotated_tc}
\end{figure}

Geometry is defined by a metric that measures length and angle in spaces, and dictates the performance of topological codes as illustrated by the rotated 2D toric code: the rotation halves the qubit count for the same number of logical qubits and code distance.   For example, for an even natural number $d$, the standard toric code is $[[2d^2,2, d]]$, while the rotated one is $[[d^2,2,d]]$.  The case $d=2$ can be visualized easily---the toric code [[8,2,2]] on the $2\times 2$ square $[0,2]\times [0,2]$ in $\bZ^2$ is rotated to [[4,2,2]] on the inscribed square with vertices $(1,0), (2,1), (1,2), (0,1)$, see Fig.~\ref{fig:small_rotated_tc}.

The 4-dimensional geometric codes live on integer lattices $\Lambda$ in the hypercubic lattice $\bZ^4$ of the 4-dimensional Euclidean space $\bR^4$.  Our codes are translation invariant, so it is better to be regarded as on a unit cell of the lattice $\Lambda$ with identified boundaries---topologically a 4-dimensional torus $T^4_\Lambda$ with a cellulation into vertices (0-cells), edges (1-cells), faces (2-cells), cubes (3-cells), and hypercubes (4-cells).  The cellulation of the 4-torus $T^4_\Lambda$ defines a (combinatorial)  geometry that we will explore for the performance enhancement of the rotated loop-only toric codes.

\subsection{4D self-correcting quantum memory}

To understand the geometry of our codes, it is helpful to have the combinatorics of the 4D cube in mind.  Fortunately, the 4D cube, also called tesseract, can be visualized as two concentric cubes  without the 4-cell---the picture for $(\interval \interval \interval \interval)_p$ in \Cref{fig:notation}.

\subsubsection{Combinatorial geometry of the tesseract}

The combinatorial geometry of the tesseract is defined by the Hamming distance.  Given an integer lattice $\Lambda$ in $\bR^4$,  a unit cell of $\Lambda$ with boundary identified is a 4-torus $T^4_\Lambda$ with a combinatorial geometry induced from the Hamming distance of $\bZ^4$.  

The combinatorics of the tesseract determine many parameters of our codes, such as the number of qubits, so it is convenient to have these numbers explicitly.  The tesseract consists of 16 vertices, 32 edges, 24 faces, 8 cubes, and 1 hypercube as can be checked in the picture for $(\interval \interval \interval \interval)_p$ in \Cref{fig:notation}.  The qubits of the loop-only toric codes live on faces of the 4-torus $T^4_\Lambda$, and the stablizers on edges and cubes.  The boundary of the tesseract is a 3-sphere consisting of 8 cubes.  If the boundary of the tesseract is identified periodically, then the resulting space is a 4-torus with a cellulation of 1 vertex, 4 edges, 6 faces, 4 cubes, and 1 hypercube.

The volume of the 4-torus $T^4_\Lambda$ is given by the determinant $\ted(L)$ of a Hermite normal form $L$, which is also the number of vertices and hypercubes of $T^4_\Lambda$.  From the combinatorics of the tesseract, it can be deduced that there are $6\times \ted(L)$ 2-cells, and $4\times \ted(L)$ edges and cubes.  It follows that the number of data qubits for the loop-only toric code on $\Lambda$ is given by $6\times \ted (L)$ with $4\times \ted (L)$ X and type Z stabilizers, respectively.  To be more convenient for syndrome extraction, $4\times \ted (L)$ ancillary qubits are used separately to measure type X and Z stabilizers.  Hence a single block of the code uses $6\times  \ted (L)$ data qubits, and $8\times \ted (L)$ ancillary qubits.  To reduce the number of physical qubits required for syndrome extraction, ancillary qubits may be reused in exchange for requiring additional time for extraction.

For each positive integer $l$, there is an integer lattice whose unit cell is the $l\times l\times l\times l$ 4-cube.  The unit cell with boundary identified is the 4-torus $T^4_l$ with volume $l^4$. The lattice unit cell has $l^4$ vertices and hypercubes, $6 l^4$ square faces, and $4 l^4$ edges and cubes.  The cells of the 4-torus $T^4_l$ form a 5-term chain complex, where each $C_i$ is the $\bF_2$-vector space generated by $i$-cells for $i=0,1,2,3,4$:

$$C_4\rightarrow C_3\rightarrow C_2 \rightarrow C_1\rightarrow C_0.$$

The homology of the 4-torus $T^4_l$ can be calculated using this chain complex.  All the 4-tori herein are homeomorphic to each other, hence all the topological information is the same independent of the lattices.  The $\bF_2$-homology of the torus is crucial for our code construction. The first homology group is $H_1(T^4,\bF_2)=\bF_2^4$, and homology classes are generated by 4 independent 1-cycles, one in each direction.  The second homology group is $H_2(T^4,\bF_2)=\bF_2^6$, which are generated by tori given by any two different directions.  Those 6 2-cycles are the support of logical operators.  One interesting technical result is to find algorithmically representatives of these 6 2-cycles in a rotated 4 torus.  One such algorithm is given below using the cup product of 1-cocyles.

\subsubsection{Weight 6 stabilizers}

The 4D loop-only toric code is part of a $(4+1)$-topological quantum field theory (TQFT), hence a 4D topological phase of matter \cite{dennis2002topological,chen2023loops}.  It follows that the representation of its mapping class group leads to topological operations.

The Hamiltonian is constructed by summing up all stablizers as local terms 
$$H=-\sum_{e}X_e-\sum_{c}Z_c,$$
where the X terms are over all edges of the 4-torus $T^4_\Lambda$, and Z terms over all cubes. Each cube has 6 faces, and each edge is touched by 6 faces in $\bZ^4$.  
Thus the Hamiltonian is a 6-body interaction,  or in coding language, all stablizers are of weight 6. 

The code space of the 4D geometric codes consists of the ground states of this Hamiltonian, and logical operators come from symmetries of this Hamiltonian.

The 4D (2,2)-toric code is called loop-only as there are no point excitations and the elementary low energy excitations are loops \cite{dennis2002topological,chen2023loops}.  This fact underpins the self-correcting and single-shot properties as codes.  The 4D loop-only toric code is proven to be a self-correcting quantum memory \cite{dennis2002topological,alicki2010thermal}. 

\subsubsection{Topological gates}\label{sec: topologicalGates}

It is a general fact that a TQFT gives rise to representations of the mapping class groups of space manifolds.  The loop-only toric code as a $(4+1)$-TQFT leads to a representation of $\textrm{SL}(4,\bZ)$, which is the mapping class group of the 4-torus $T^4$ \cite{dennis2002topological,chen2023loops}.

It follows that there are also topological gates from the representation of the mapping class group $\textrm{SL}(4,\bZ)$, which is generated by two matrices \cite{trott1962pair}:

$$s=
\begin{pmatrix}
0 & 1 & 0 & 0\\
0 & 0 & 1 & 0\\
0 & 0 & 0 & 1\\
1 & 0 & 0 & 0\\
\end{pmatrix},
t=
\begin{pmatrix}
1 & 0 & 0 & 0\\
1 & 1 & 0 & 0\\
0 & 0 & 1 & 0\\
0 & 0 & 0 & 1\\
\end{pmatrix}.
$$

Let $\{e_i\}_{i=0}^3$ be the standard row basis of $\bR^4$ with all coordinates=0 except $1$ at the $(i+1)$-th place.  Then $s$ sends $e_i$ to $e_{i-1}$ with indices mod 4, and $t$ sends $e_i$ to $e_i$ if $i\neq 1$ and $e_1$ to $e_{0}+e_{1}$.

Let $T^4$ be the 4-torus from identifying the boundaries of the tesseract.  Then the matrix $L$ of a 4D lattice induces an automorphism of the first homology $H_1(T^4,\bZ)$ with the standard 1-cycles from basis $\{e_i\}$.  The squares in the $(i,j)$-plane for two different $i,j$ in $\{0,1,2,3\}$ are identified to 6 2-tori $\{e^2_{i,j}, ij=01,02,03,12,13,23\}$, which form a basis of $H_2(T^4,\bZ)$.

The action of the mapping classes $s,t$ on first homology $H_1(T^4,\bZ)$ induces an action on $H_2(T^4,\bZ)$, leading to topological gates $S,T$ in some basis:
$$S=\textrm{Permutation gate of}\;\;((01)(03)(23)(12))((02)(13))$$
$$T=CX((12),(02)) \;\; CX((13),(03)).$$
Since those are topological gates,  they are therefore also gates for all 4D geometric codes in appropriate bases.  

Logical gates of topological codes come from symmetries of their Hamiltonians, and the ones that we focus on later are space group symmetries and ZX dualities.  The more subtle topological symmetries also lead to gates of the logical qubits.
If those topological gates can be implemented efficiently, then they could be included into the logical gate set, hence a useful part of the Clifford synthesis.  Therefore, it would be interesting to find local low-depth implementations of those topological gates.

\subsection{Geometry change by rotation}

The 2D toric code was invented in Ref.~\cite{kitaev1997quantum}, and its rotation first appeared as the Wen plaquette model in physics \cite{wen2003quantum}, though its significance for error correction was not explored until later.

The geometry of the standard lattices with unit cells 4-cube $l\times l\times l\times l$ comes from the Manhattan distance of $\bZ^4$.  The induced geometry of the torus $T^4_\Lambda$ from the Manhattan distance for a general lattice $\Lambda$ can be very different from the usual one.  In the example in \Cref{geometric} for $d=2$, each essential 1-cycle has length=2 in this induced geometry from Manhattan distance, but their standard Euclidean length is $\sqrt{2}$.  The preservation of code distance by rotation while reducing the area/volume, hence qubit count, is the key to the performance enhancement of topological codes by geometry. 

\subsubsection{Hermite normal form of 4D lattices}

An integer lattice $\Lambda$ in $\mathbb{R}^4$ is a collection of periodic points with integral coordinates.  The lattice $\Lambda$ can be represented by an integral matrix $L$ that consists of 4 rows as the 4 generating vectors of the lattice.  Many different such generating matrices represent the same lattice $\Lambda$: two such matrices represent the same lattice if and only if they are related by a multiplication from the left by a matrix in $\textrm{SL}(4,\bZ)$ and/or from the right by a signed permutation matrix.

Using the two equivalences, any such generating matrix $L$ for a lattice can be represented by a matrix $L$ in the Hermite normal form (HNF), which satisfies the following conditions: $L=(a_{ij})$ is an upper triangular matrix such that the diagonal entry in each column is the largest in the column, i.e.

\begin{align}
    i>j \; \Longrightarrow   a_{ij}=0,
\end{align}
\begin{align}
    i<j \; \Longrightarrow 0\leq  a_{ij} < a_{jj}.
\end{align}

Three lattices that we are most interested in have the following HNFs:

\begin{align}
L_1 &=
\begin{pmatrix}
1 & 1 & 1 & 1\\
1 & -1 & 1 & -1\\\
1 & 1 & -1 & -1\\
1 & -1 & -1 & 1
\end{pmatrix} \sim 
\begin{pmatrix}
1 & 1 & 1 & 1\\
0 & 2 & 0 & 2\\\
0 & 0 & 2 & 2\\
0 & 0 & 0 & 4
\end{pmatrix}, \label{eq:hadamard-lattice-hnf}\\
L_2 &= \begin{pmatrix}
1 & 0 & 0 & 3\\
0 & 1 & 0 & 5\\
0 & 0 & 1 & 7\\
0 & 0 & 0 & 16\\
\end{pmatrix},\\
L_3&= \begin{pmatrix}
1 & 0 & 1 & 6\\
0 & 1 & 0 & 11\\
0 & 0 & 3 & 9\\
0 & 0 & 0 & 15\\
\end{pmatrix}.
\end{align}

We will refer to the first lattice as the Hadamard lattice, as the matrix $L_1$ is a Hadamard matrix $H\otimes H$, where
$
H=\left(
\begin{array}{cr}
1 & 1 \\
1 & -1\\
\end{array}
\right).
$
The Hadamard lattice has determinant 16.  We will refer to the second lattice by $\ted16$, and to the third by $\ted45$.  

\subsubsection{4D loop-only toric codes on general lattices}

In general, it can be difficult to calculate the distance of a code.  The geometry of the torus from celluations leads to good estimates of the code distance, and algorithms for its calculation.

Essential cycles or cocycles are important for computing code distance and for decoding topological codes.  These are cycles that are not boundaries.  Code distance is closely related to the combinatorial length/area/volume of essential cycles or cocycles.  In general the code distance is given by the combinatorial systole or co-systole of the cell complex \cite{freedman2002z2}.  

For the loop-only toric code, the code distances $d_X$ and $d_Z$ are given as:  $d_X$ is the minimal area of essential 2-cycles, while $d_Z$ is the minimal area of the essential 2-cocycles.  Then the code distance is the minimum of the two.

The Hadamard code has distance 8 and the Det45 code has distance 15 \cite{aasen2025topological}.
In \Cref{tab:code_parameters} we list various 4D codes and their corresponding lattices and distances. 

\begin{table}\label{table: codes}
    \centering
    \begin{tabular}{ccccccc cccccccc}
            \toprule
            \textbf{Name}& \textbf{Det} &\textbf{n} &\textbf{k} &\textbf{d} &  $a_{11}$&$a_{12}$&$a_{13}$&$a_{14}$&$a_{22}$ &$a_{23}$&$a_{24}$&$a_{33}$&$a_{34}$&$a_{44}$\\
            \midrule
            Det2 &2 & 12 & 6 & 2 & 1 & 0& 0& 1& 1& 0& 1& 1& 0 &2\\
            Det3 &3 & 18 & 6 & 3 & 1 &0& 0& 1& 1& 0& 1& 1& 1& 3\\
            Det5 &5 & 30 & 6 & 4 & 1 &0& 0& 1& 1& 0& 2& 1& 3& 5\\
            Det9 &9 & 54 & 6 & 6 & 1 &0& 0& 5& 1& 0& 6& 1& 7& 9\\ 
            Hadamard &16 & 96 & 6 & 8 & 1 &1& 1& 1& 2& 0& 2& 2& 2& 4\\
            Det16 &16 & 96 & 6 & 8 & 1 &0 & 0& 3& 1& 0& 5& 1& 7& 16\\
            Det18 &18 & 108 & 6 & 9 & 1 &0& 0& 3& 1 &0 &5 &1& 7& 18\\
            Det45 &45 & 270 & 6 & 15 & 1 &0& 1& 6& 1& 0& 11& 3& 9& 15\\
            Det68 &68 & 408 & 6 & $\leq $18 & 1& 0 &0 &21& 1& 1 &24& 2 &30 &34\\
            Det152 &152 & 912 & 6 & $\leq $30 &  1& 0& 0& 115& 1 &0 &124& 1 &136 &152\\
            \bottomrule
        \end{tabular}
    \caption{A list of codes with optimal code parameters for a given determinant. 
    We have only listed the non-trivial entries of the lattice $L$ in Hermite normal form following the convention $L=(a_{ij})$ as described in the main text.
    Exact upper bounds for the $\ted68$ and $\ted152$ codes are computed from probabilistic methods~\cite{leon1988}.
    }
    \label{tab:code_parameters}
\end{table}

\subsubsection{Optimal lattice conjecture}\label{sec: optimal lattice conjecture}

The standard 4D loop-only toric code on $l\times l\times l\times l$ lattice has code parameters $$[[6l^4, 6, l^2]] = [[6d^2, 6, d]],$$ while the optimal rotated loop-only toric codes are given by 
$$[[\alpha(d), 6, d]]$$
for some integer valued function $\alpha(d)$. 
It is natural to ask what is the minimum of $\alpha(d)/d^2$ over all distances, and secondly what does $\alpha(d)/d^2$ limit to as $d \rightarrow \infty$.
The Det45 code has $\alpha(d)/d^2 = 1.2$, hence providing an upper bound on the minimum of $\alpha(d)/d^2$. 
We conjecture that $\alpha(d)/d^2 \rightarrow 1$ in the limit of $d \rightarrow \infty$.
The significant savings of at least a 5-fold reduction in the number of physical qubits plus the single-shot property make rotated toric codes attractive candidates for current and near-term hardware implementation. 

\subsubsection{Logical operators}\label{logical basis}

Given a lattice $\Lambda$ with an HNF $L$, the code subspace of the 4D geometric code $C_\Lambda$ is spanned by the second homology $H_2(T^4_\Lambda, \bF_2)$ of the 4-torus $T^4_\Lambda$, so there are always 6 logical qubits in a 4D geometric code. To find logical operators explicitly, logical qubits need to be given in some explicit basis.  The common practice is to use the basis given by a collecting of independent maximally commuting logical operators: a basis vector is determined by the $\{\pm 1\}$ eigenvalues of each logical operator.  

To find a commuting independent set of 6 logical operators, we need to find either 6 2-cycles or 6 2-cocyles in the 2-skeleton of the lattice $\Lambda$.  The 2-cycles and 2-cocyles  serve as the supports of the logical $X$ and $Z$ operators.   Hence general methods are needed to produce such explicit representatives.  Below we describe explicit algorithms to generate such 2-cycles and 2-cocycles.

There are many practical ways to find such representatives.  For small determinant codes, a direct calculation from the HNF might be the most straightforward.  But when the determinant is large, this direct approach becomes less practical and computing expensive.

When the determinant of a code is odd, there is a direct way to write down 2-cycle representatives.  Take any two directions of the 4 coordinate directions such as $x,y$, the sum of all 2-cells in the parallel planes of these two directions is a 2-cycle.  But when the determinant is even, this sum might be the 0-cycle, so another general method is needed.

Given an HNF $L$, it is easy to write down 4 independent 1-cycle representatives in the edges of the lattice of $\Lambda$ using the 4-rows of $L$, denoted as $c_i,i=1,2,3,4$. 

The Poincare dual of a 1-cycle $c_i$ is a 3-cocyle represented by cubes in the dual lattice of $\Lambda$. By shifting these 3-cocyles from the dual lattice to the direct lattice $\Lambda$, we obtain 4 3-cocycles in the direct lattice, which form a basis of $H^3(T^4_\Lambda, \bF_2)$.  The intersection of such a pair is a representative of a 2-cocyle in $H^2(T^4_\Lambda, \bF_2)$ given by 2-cells of the direct lattice \cite{chen2023higher}.

More directly explicit, first use the HNF $L$ of $\Lambda$ to find 4 independent 1-cocycles in the edges of $\Lambda$ for $H^1(T^4_\Lambda, \bF_2)$. These can be done either by direct calculation or using duality in the 3-skeleton of the 4-torus similar to above.  

Let $\alpha$ and $\beta$ be two such 1-cocyles in the edges of $\Lambda$, then their cup product can be calculated by the following formula \cite{chen2023higher,chen2023loops}:

$$ \alpha\cup \beta (\interval \interval)=\alpha(\interval 0)\beta (1\interval )+\alpha(\interval 0)\beta (\interval 1).$$

The meaning for the formula is as follows: without any loss of generality, we may assume that any square (2-cell) $e^2$ in the lattice $\Lambda$ is identified with the standard square in the $x-y$ plane.
On the boundary of the standard square, there are two paths of two edges from the origin $(0,0)$ to $(1,1)$: $(\interval 0)+\beta (1\interval )$ and $(\interval 0)+(\interval 1)$. Then $\alpha\cup \beta (\interval \interval)$ is the sum of the evaluations of the $\alpha$ and $\beta$, respectively.

From the evaluations of $\alpha\cup \beta (\interval \interval)$ on each 2-cell, we obtain a 2-cocyle $\alpha \cup \beta \in Z^2(T^4_\Lambda, \bF_2): C_2(T^4_\Lambda)\rightarrow \bF_2$ by summing up delta functions on the 2-cells with evaluation=1.

To get 2-cycles in $\Lambda$, we dualize the 2-cocyles $\alpha \cup \beta$ obtained from above.

\subsection{Single-shot}

The origin of single-shot error correction can be traced back to the generalization of the self-correcting quantum memory of the 4D loop-only toric code \cite{dennis2002topological} to a self-correcting quantum computer \cite{bombin2013self}.  The original single-shot theory is formulated for topological quantum subsystem codes with stochastic local excitation errors \cite{bombin2015single}.  A general theory of single-shot error correction for adversarial noise only is in Ref.~\cite{campbell2019theory}, where it was shown that every code can be made single-shot with non-local checks and action on many qubits.  We will limit our discussion to topological CSS quantum codes with local excitation errors and local measurements in this section.  This restriction to topological CSS codes is not necessarily serious as most CSS codes are topological \cite{freedman2021building}.  

Syndrome extraction by measuring the stablizers is crucial for error detection, decoding, and correction.  To have high confidence of the measurement results, it is in general necessary to repeat local measurements $f(d)$ rounds for a code with distance $d$, where $f(d)$ is some function of $d$.  For codes that are not single-shot, the function $f(d)$ grows as $d$ increases, though the best function form seems to be unknown.  Extraction of the syndromes
in quantum computing, which can make several physical operations and measurements to achieve, is in general slow and faulty, so only a single round or a constant few of local measurements in a single-shot code is a significant advantage for hardware implementation.  Further, the connectivity of those syndromes and excitations is important for being self correcting.  Single-shot for topological codes is a property of the codes due to it being self-correcting, and then in turn maintaining a low depth for that single shot.
Single-shot error correction for topological quantum codes in \cite{bombin2015single} is an important alternative for decoding noisy syndrome extraction data.  

Given a topological CSS quantum code as the ground-state space of a local Hamiltonian:
$$H=-J\sum_{s\in S}s,$$
where $S$ is a set of local stablizer generators.  Then all errors would be Pauli, and are the same as excitations of this Hamiltonian.  More precisely, a topological quantum code consists of a family of such codes with different system sizes $n$, which usually is the number of $k$-cells in a lattice for some $k$.  

Let $\mathcal{P}_n$ be the Pauli group of $n$-qubits.  A Pauli error $E\in \mathcal{P}_n$ has a unique decomposition $E=D\cdot L\cdot s$, where $D$ is the de-stablizer, $L$ the logical part, and $s$ the stablizer part.  The de-stablizer $D$ is determined by the error syndrome.  A successful decoding would find a recovery logical part $L'$ modulo the stablizers from the error syndrome, i.e., $E=D\cdot L'\cdot s'$ such that $s'$ is also a stablizer. For every error syndrome $\sigma$, we choose an operator $D_\sigma$ that takes the states back to the code subspace.

Given a Pauli noise channel $\cE=\{\sqrt{p_\cE(E)}E\}_{E\in \mathcal{P}_n}$, its ideal error correction failure probability is defined by 
$$\textrm{fail}(\cE)=\sum_{L,L\neq 1}\sum_{\sigma}p_{\cE}(D_\sigma L).$$ The syndrome distribution of $\cE$ is defined as
$$q_\cE(\sigma)=\sum_{L}p_\cE(D_\sigma L).$$

For single-shot error correction, we first measure locally all stablizer generators in $S$.  Instead of the perfect syndromes $\sigma$, we assume that the resulting syndromes are $\sigma+\omega$ with some probability distribution $r(\omega)$ independent of $\sigma$.  Next find a $\sigma_0$ with minimal cardinality so that $\sigma+\omega+\omega_0$ are valid syndromes. Then the single-shot error correction is done by applying $D_{\sigma+\omega+\omega_0}$.

Given two real numbers $\epsilon>0,\tau>0$, an $(\epsilon,\tau)$-Pauli excitation error channel $\mathcal{E}_{\epsilon,\tau}$ is a quantum channel such that 
$$ \text{fail}(\mathcal{E}_{\epsilon,\tau})\leq \epsilon, ||q_{\mathcal{E}_{\epsilon,\tau}}||\leq \tau,$$ where $\epsilon$ is some measure of logical fidelity, and $\tau$ characterizing the accumulation of physical errors. 
Let $\cN_{\epsilon,\tau}$ be the set of all $(\epsilon,\tau)$-Pauli excitation error channels $\mathcal{E}_{\epsilon,\tau}$ of a topological CSS code.

For any $\eta >0$, let $\mathcal{R}_\eta$ be the set of all recovering operations $\mathcal{R}$ such that 
$$q_{\cR}(\omega')=\sum_{\omega:\omega+\omega_0=\omega'}r(\omega')$$
for some $\eta$-bounded probability distribution $r(\omega)$.

\begin{definition}
A topological CSS quantum code is single-shot if there exist two system-size-independent real constants $\tau_0>0, \eta_0>0$ so that for any real numbers $\eta>0, \tau>0, \tau'=\frac{(\frac{\eta}{\eta_0})^{1/2}}{1-(\frac{\eta}{\eta_0})^{1/2}}$ such that $\eta < \eta_0, \tau+\tau'< \tau_0$, the recovery operations $\mathcal{R}_\eta$ and error channels $\cN_{\epsilon,\tau}$ satisfy the following set-composition equation: 
\begin{equation}
    \mathcal{R}_\eta \cdot \mathcal{N}_{\epsilon,\tau} \subseteq \mathcal{N}_{\epsilon+\delta,\tau'} \label{eq:single}
\end{equation}
when restricted to the code subspace, where the positive real functions $\epsilon=\epsilon(n,\tau),\delta=\delta(n,\tau, \eta, |S|)$ both go to $0$ as the system size $n$ goes to $\infty$ with  $\tau+\tau' <\tau_0$. Note that $\tau'$ goes to $0$ when $\eta$ goes to $0$. 
\end{definition}

Equation~\ref{eq:single} tells us that error correction after one round below certain critical values removes errors from the system while keeping the encoded information as intact as possible.  But the noise in the recovery operation still leaves behind residual errors in the system.  Both self-correcting and single-shot for topological quantum codes are related to confinement in topological phases of matter \cite{bombin2015single}.  

By Theorem 14 of \cite{bombin2015single}, self-correcting topological CSS codes are single-shot.  The 4D loop-only toric code is proven to be self-correcting \cite{alicki2010thermal}, hence it is provably single-shot.

A topological quantum code is self-correcting if it retains a non-trivial topological order at non-zero finite temperature.  So in the thermodynamic limit, the quantum memory is not destroyed at non-zero finite temperature without any active error correction. 

Self-correcting of topological CSS quantum codes is a manifestation of the connectivity structures of excitation and syndrome graphs \cite{bombin2013self}.  The connectivity correspondence between excitation and syndrome graphs leads to local structures in the Pauli excitation error models for topological quantum codes.  Then the single-shot property follows from graph combinatorics of the self-correcting codes with explicit constants \cite{bombin2015single}.

Concretely, the single-shot property of 4D geometric codes can be seen from the redundancy of X and Z stablizers encoded in the 0- and 4-cells in the 5-term chain complex of the celluations of the 4-tori:
\begin{equation}
\begin{tikzcd}
C_4 \arrow[r, bend left=20, "\partial_4"] & 
C_3 
\arrow[l, bend left=20, "\partial^T_4"]
\arrow[r, bend left=20, "\partial_3"] &
C_2
\arrow[l, bend left=20, "\partial^T_3"]
\arrow[r, bend left=20, "\partial_2"] &
C_1 
\arrow[l, bend left=20, "\partial^T_2"]
\arrow[r, bend left=20, "\partial_1"]&
C_0
\arrow[l, bend left=20, "\partial^T_1"]
.
\end{tikzcd}
\end{equation}

Single-shot makes it possible to design fault-tolerant schemes such that all elementary logical operations can be performed in constant time.  After a round of error correction in a single-shot code, there could still be residual local noise, but the errors are kept low so that they do not accumulate enough to derail a computation.  The single-shot property of a code neither implies perfect decoding with a single round of syndrome extraction nor requires the use of only a single round of error syndrome extraction.

\subsection{Decoders}
In our numerical studies we use a modified belief-propagation decoder and a decoder similar to an optimized look-up table.

\subsubsection{BP+OSD decoder}\label{sec:bp-osd}
The {\em belief-propagation ordered statistics decoder}~\cite{panteleev2021degenerate,roffe2020decoding} has become the {\em de facto} standard for decoding of qLDPC codes, so we use the BP+OSD decoder in our simulation for ease of comparison with other codes and decoders. We essentially mimic the decoder setup from Ref.~\cite{bravyi2024high} using the open-source LDPC library for the BP+OSD implementation~\cite{roffe2022}, and use it to jointly decode a window of $d$ rounds of noisy syndrome extraction.  Given the single shot nature of the code, we expect a constant window size (i.e., independent of $d$) will suffice to achieve similar performance, but use $d$ rounds here to facilitate comparison to the results in Ref.~\cite{bravyi2024high}.

\subsubsection{Power decoder}\label{sec:power-decoder}
We also use a decoder based on the algorithm described in Ref.~\cite{aasen2025topological}.
Given a target syndrome $t$, the algorithm attempts to find a subset from a known set $S$ of syndromes that matches $t$.  Subsets of progressively larger size are checked until either a match is found or a size limit is reached.  Since the search space is the powerset of $S$, we call this decoder the \emph{power decoder}.
The algorithm is detailed in~\Cref{power-decoder}.

Brute force enumeration over the powerset of syndromes $S$ is prohibitively costly.  The set of unique syndromes due to single $X$ errors is typically in the hundreds for our circuits.  We use a few tricks to manage the run time.  First, we set a limit $k_\text{max}$ on the solution size.  If each element of $S$ occurs independently with low probability, then the probability of needing many such elements is negligible beyond a certain size.  
Second, with a limit on solution size, we can then prune smaller subset candidates for which a solution is not possible.  If $w_\text{per}$ is the maximum Hamming weight of any individual syndrome, then the maximum weight of any combination of $k$ syndromes is at most $k\cdot w_\text{per}$.  We can, therefore, reject partial solutions for which the remaining target syndrome has high weight.
Finally, we pre-compute a lookup table of combinations of syndromes, up to some fixed size.  This technique is similar to a ``meet in the middle'' approach.  We enumerate over the powerset to find part of the solution, and then search the table for the other part.
This reduces the value of $k_\text{max}$ necessary to find a solution.

There is freedom in choosing the initial set of syndromes $S$ and corresponding Pauli errors $E$.
For a given syndrome extraction circuit, we enumerate over all possible single faults. Each fault induces a syndrome and a residual error on the output qubits.  The most straightforward approach then is to use those syndromes and residual errors directly.
In general, however, there can be several faults that induce the same syndrome but distinct residual errors.  We therefore take a different approach.

We partition faults into four sets: qubit faults, measurement flips, ancillary faults, and partial faults.  The qubits faults are the set of single-qubit errors on code block qubits, which occur just prior to a syndrome extraction round.  Each fault has a unique syndrome and residual error.  The measurement flips are faults that flip a syndrome measurement outcome, but otherwise have no residual error.  Ancillary faults are those that occur on an ancilla qubit and cause (high weight) residual errors on code block qubits.  Finally, partial faults are the set of remaining faults that occur part way through a syndrome extraction circuit.  We preferentially use the syndromes and residual errors of the qubit and measurement flip faults.  Further, we find that using a trivial recovery for partial faults, instead of using the residual error directly, yields better results.

\begin{algorithm}
\caption{\label{power-decoder}Power decoder}
\begin{algorithmic}[1]
\REQUIRE Syndromes $S=\{s_1,\ldots,s_n\}$, Pauli errors $E=\{e_1,\ldots,e_n\}$, lookup table $L$, integer $k_{\text{max}}$ target syndrome $t$.
\STATE $I'\leftarrow \emptyset$, $t_{\text{best}} \leftarrow t$
\STATE $w_{\text{per}} \leftarrow \max \{\text{wt}(s)~\forall s\in S\}$
\STATE $w_{\text{table}} \leftarrow \max \{\text{wt}(L[s])~\forall s\in L\}$
\FOR {$k$ in $0$ \TO $k_{\max}$}
    \STATE $w_{\text{max}} \leftarrow (k-1)w_{\text{per}} + w_\text{table}$
    \FOR {$I$ in PrunedSubsets($S$, $k$, $w_{\text{per}}$, $w_\text{max}$, $t$)}
    \STATE $t' \leftarrow t + \sum_{i \in I} s_i$
    \IF {$t'\in L$}
        \RETURN $\prod_{i \in I \cup L[t']} e_i$
    \ENDIF
    \IF {$\text{wt}(t') < \text{wt}(t_{\text{best}})$}
        \STATE $I' \leftarrow I$, $t_{\text{best}} \leftarrow t'$
    \ENDIF
    \ENDFOR
\ENDFOR
\RETURN $\prod_{i\in I'} e_i$
\end{algorithmic}
\end{algorithm}

\begin{algorithm}
\caption{\label{pruned-subsets}PrunedSubsets}
\begin{algorithmic}[1]
\REQUIRE Syndromes $S=\{s_1,\ldots,s_n\}$, integers $k$, $w_{\text{per}}$, $w_\text{max}$ target syndrome $t$.
\IF {$k == 0$}
\STATE \textbf{yield} $\emptyset$
\ELSE
\FOR {$i \in [1,n-w+1]$}
    \STATE $t' \leftarrow s_i + t$
    \IF {wt($t'$) $\le w_\text{max}$}
        \FOR {$I \in$ PrunedSubsets$([s_{i+1},\ldots,s_n], k-1, w_{\text{per}}, w_\text{max}-w_\text{per}, t')$}
             \STATE \textbf{yield} $\{s_i\} \cup I$
        \ENDFOR
        
    \ENDIF
\ENDFOR
\ENDIF
\end{algorithmic}
\end{algorithm}

\section{Performance and comparison}

\subsection{Syndrome extraction circuits}
We consider two types of syndrome extraction circuits which may be useful in different settings \cite{aasen2025topological}. 
The {\em compact circuit} (Circuit~\ref{compactse}) has short overall depth and is well suited for machine architectures with a high degree of parallelism. 
The {\em starfish circuit} (Circuit~\ref{starfish}) has depth twice as large, but propagates error chains in a favorable way yielding improved performance, and may be well suited for architectures which are throttled by gate parallelism. 
Moreover, as the $X$ and $Z$ syndrome extraction circuits are implemented sequentially, it is well suited for  ancilla reuse during extraction and qubit-limited platforms. 
In \Cref{app:secircuits} we write out each syndrome extraction circuit explicitly.
The CNOT gates in the syndrome extraction circuits are organized according to the four cardinal directions in four dimensions, and executed in parallel.

\subsection{Memory performance}\label{sec:memory-performance}

It is instructive to evaluate the performance of encoded memory or logical idle operation using various instances of the 4D geometric code under circuit-level depolarizing noise. We consider a benchmark where $d$ rounds of noisy syndrome extraction are performed for a distance $d$ code, followed by a round of noiseless syndrome extraction. We estimate the probability $p_{\text{fail}}$ that there is a logical failure in this $d$ round circuit, and report $\frac{p_{\text{fail}}}{d}$ as the {\em logical failure probability per round}. Unsurprisingly, the decoder choice can have a large impact on performance, and we consider two options: (i) separate decoding of each syndrome extraction round (``single-shot decoding'') using the power decoder from \Cref{sec:power-decoder}, and (ii) joint decoding of all $d$ rounds using BP+OSD~\cite{panteleev2021degenerate}. We treat these two options as (informal) extreme cases, in the sense the power decoder is too slow to jointly decode $d$ syndrome rounds of the larger code instances we consider here, while BP+OSD has low accuracy in the single-shot setting. In all cases we leverage the CSS structure of the code, and decode $X$ and $Z$ errors separately.

\begin{figure}[h!t]
    \vspace{-20pt}
    \centering
    \begin{subfigure}[b]{0.65\textwidth}
        \centering
        \includegraphics[width=\textwidth]{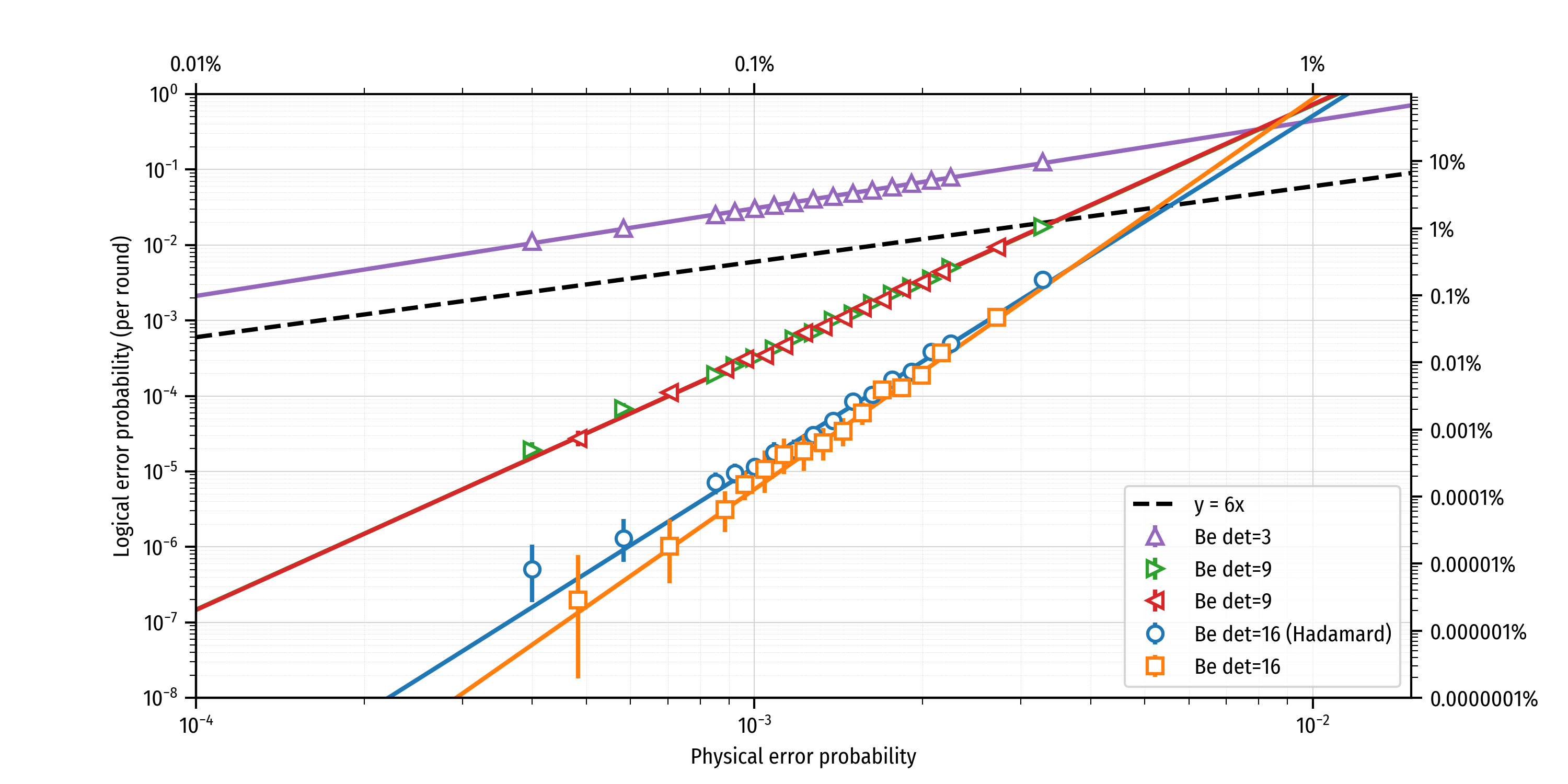}
        \caption{Single-shot power decoding of 1 noisy syndrome extraction round.}
        \label{fig:be-power-singleshot-single-round}
    \end{subfigure}
    \begin{subfigure}[b]{0.65\textwidth}
        \centering
        \includegraphics[width=\textwidth]{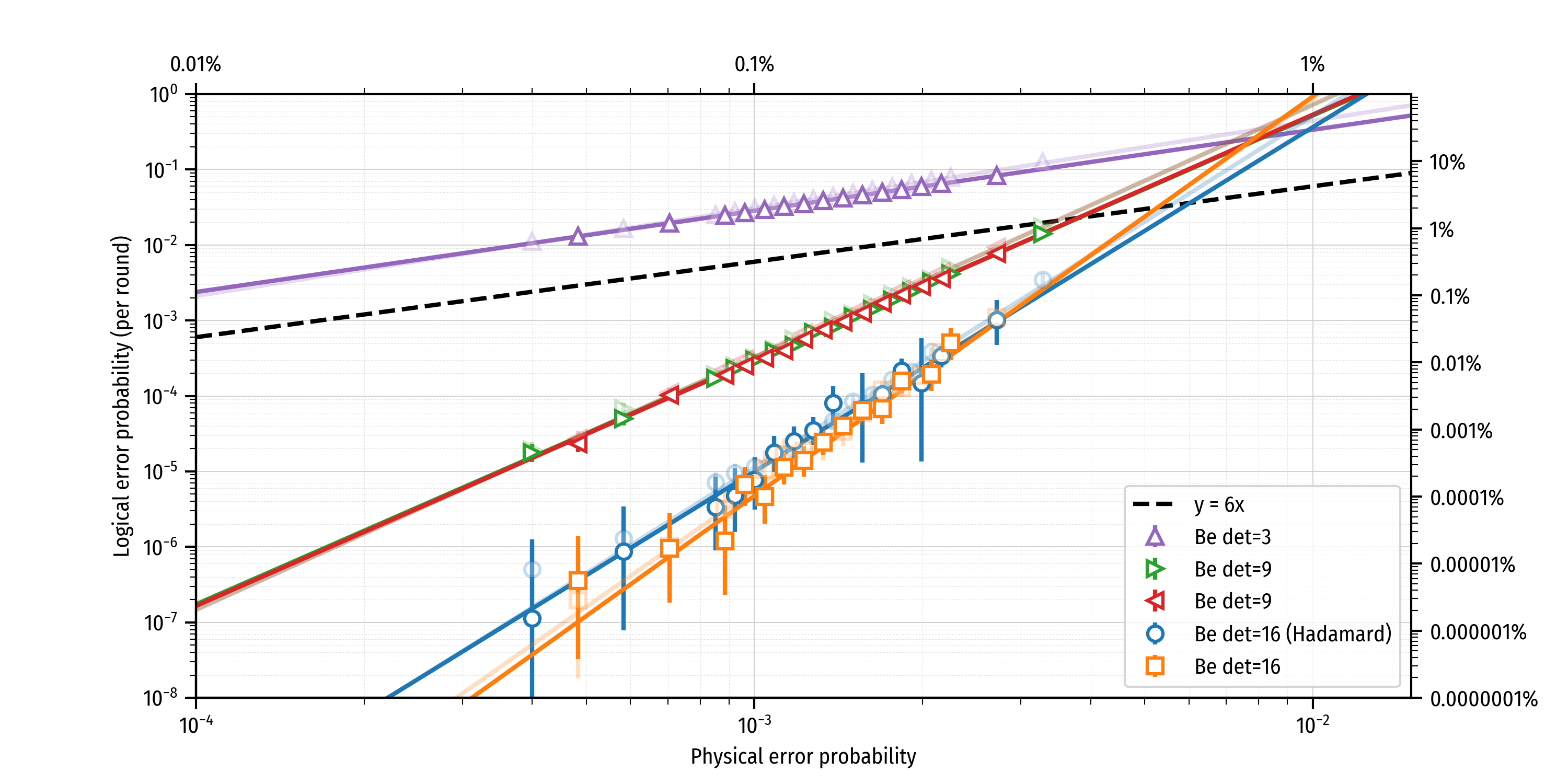}
        \caption{Single-shot power decoding of $d$ noisy syndrome extraction rounds.}
        \label{fig:be-power-singleshot}
    \end{subfigure}\par
    \begin{subfigure}[c]{0.65\textwidth}
        \centering
        \includegraphics[width=\textwidth]{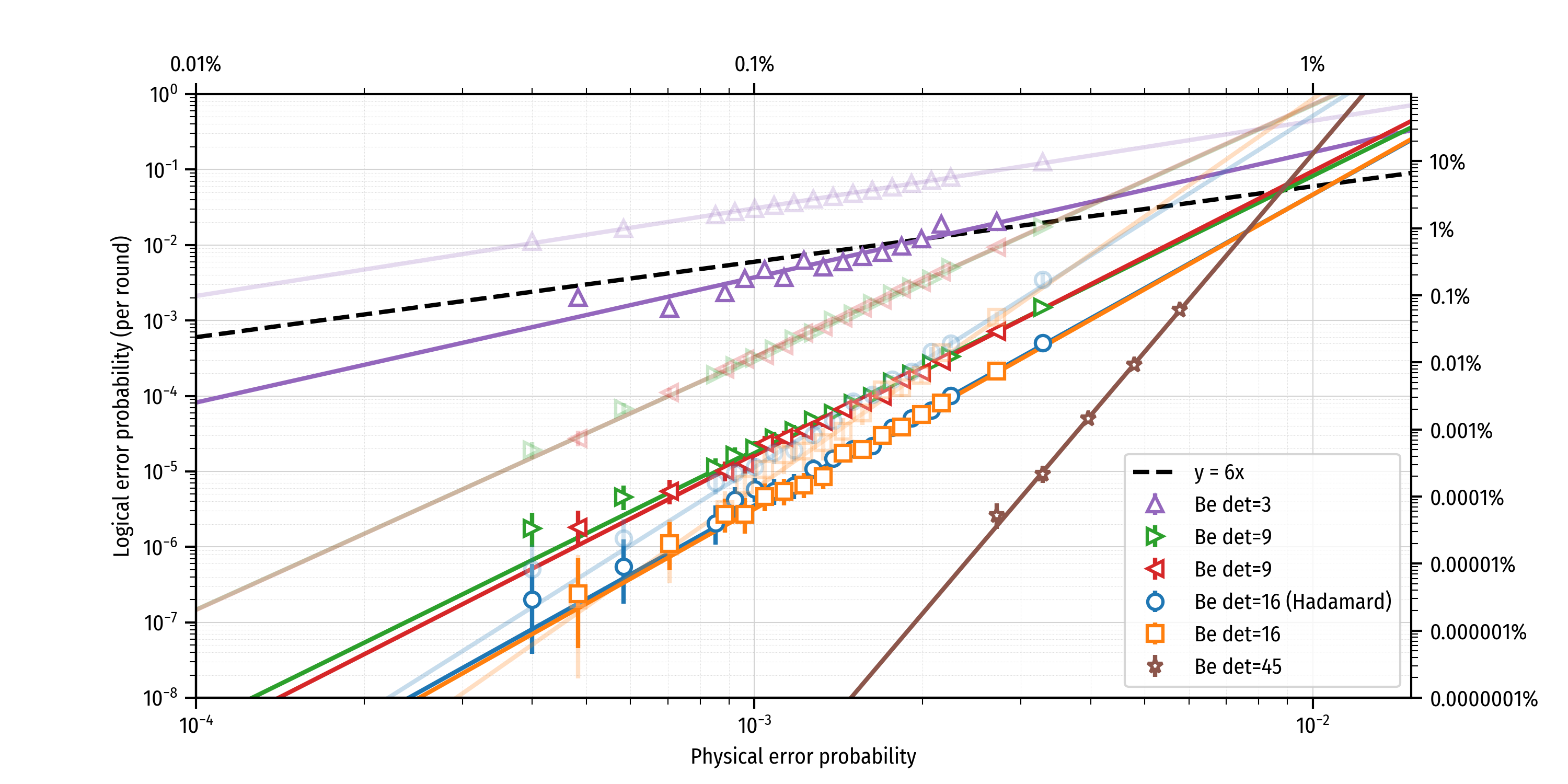}
        \caption{Joint BP+OSD decoding of $d$ noisy syndrome extraction rounds.}
        \label{fig:be-bposd-monolithic}
    \end{subfigure}
    \captionsetup{font=footnotesize, skip=2pt}
    \caption{Encoded memory performance under circuit-level depolarizing noise for various 4D geometric code instances. We simulate (a) a single round of noisy syndrome extraction with the power decoder, (b) $d$ rounds of noisy syndrome extraction with single-shot power decoding, and (c) with joint multi-round BP+OSD decoding. The single-shot, single-round results are shown in all plots for ease of comparison. Detailed code parameters can be found in \Cref{app:simulation-code-parameters}. Log-linear fits to the observed logical error rates are consistent with the simulations achieving code distance except to determinant 3 with single-shot decoding, and determinant 16 with joint multi-round BP+OSD decoding. The logical performance should be compared to 6 unencoded qubits (black dashed line), since all instances of the 4D geometric code encode 6 qubits.}
    \label{fig:memory-simulations}
\end{figure}

Using this benchmark we consider instances of the 4D geometric code with HNF determinants 3, 9, and 16, and corresponding distances 3, 6, and 8 (detailed code parameters can be found in \Cref{app:simulation-code-parameters}). For detereminants 9 and 16 we consider two lattices with different HNF but same distance. We choose gate orderings that are favorable to each decoding strategy, namely, starfish ordering for single-shot power decoding, and compact circuits for joint multi-round BP+OSD decoding (See \Cref{app:secircuits} for details). The results are illustrated in \Cref{fig:memory-simulations}. Notably, both decoding strategies point to a threshold approaching 1\%, with single-shot power decoding approaching 0.8\% and multi-round BP+OSD decoding approaching 1.5\%. 

The single-shot power decoding of a single noisy syndrome extraction round (\Cref{fig:be-power-singleshot-single-round}) yields performance that is consistent with single-shot power decoding of a sequence of $d$ rounds of syndrome extraction (\Cref{fig:be-power-singleshot}), where $d$ is the distance of the code. This implies that the decoding statistics and behavior for each decoding problem are, to good approximation, independent even if they do not satisfy strict independence requirements~\footnote{Exhaustive enumeration allows us to confirm that there are violations of strict independence as defined by Ref.~\cite{cross}. However, these are very high order, low probability error events.}. Our single-shot power decoding strategy fails to achieve code distance for $d=3$, but log-linear fits indicate distance is achieved for the larger lattices, with pseudo-thresholds approaching 0.4\%, and 0.6\% for distance 6 and 8 (the observed logical failures are also consistent with code distance being achieved). 

The joint BP+OSD decoding strategy shows results (\Cref{fig:be-bposd-monolithic}) consistent with distance being achieved for the determinant 3 and 9 codes, but the determinant 16 data has behavior consistent with distance 6 instead of 8. The logical error rates are lower than the single-shot power decoder strategy and have larger pseudo threshold, approaching 0.2\%, 0.9\%, and 1\% for determinants 3, 9, and 16 respectively.

\subsection{Performance enhancement}

The single-shot property benefits platforms like neutral atoms and trapped ions by shortening the logical cycle. Recent experiments and proposals for long distance entangling gates ~\cite{wang2025demonstration,bravyi2024high} mean that 4D geometric codes can also be realized by superconducting qubits.

The performance of a code depends strongly on the noise model of physical devices.  Pre-/post-selection could be used to enhance performance of codes, especially when the underlying noise model includes loss or leakage errors. 

If we have a logical error rate $\epsilon$, and $\frac{1}{\epsilon}$ gates are performed, then there is a non-negligible chance that at least one gate will have an error, possibly causing the overall output to be in error.  However, in many cases, we can detect in the decoding process that it is likely for an error to occur.  If the weight of correction that we need to apply is half the code distance, for example, a logical error becomes likely.  Thus, it may be useful to postselect out all runs in which such a large correction needs to be applied, while still correcting runs with only small corrections.  This will reduce the probability of an undetected logical error, while leading to a manageable post-selection rate. 

\subsection{Code comparisons}

While 4D geometric codes are conceptually interesting, their good performance also places them among implementable codes on current hardware.  Apple-to-apple code performance comparison is tricky.  For performance comparison, we will focus on the simulated logical error rate per logical qubit per error correction round at the same fixed physical error rate and the pseudo-threshold \cite{svore2005flow} reported in the literature.  Since logical error rates are exponentially senstive to the code distance, we choose to compare codes with similar distances and ballpark encoding rates.

\begin{figure}
    \centering
    \includegraphics[width=0.5\linewidth]{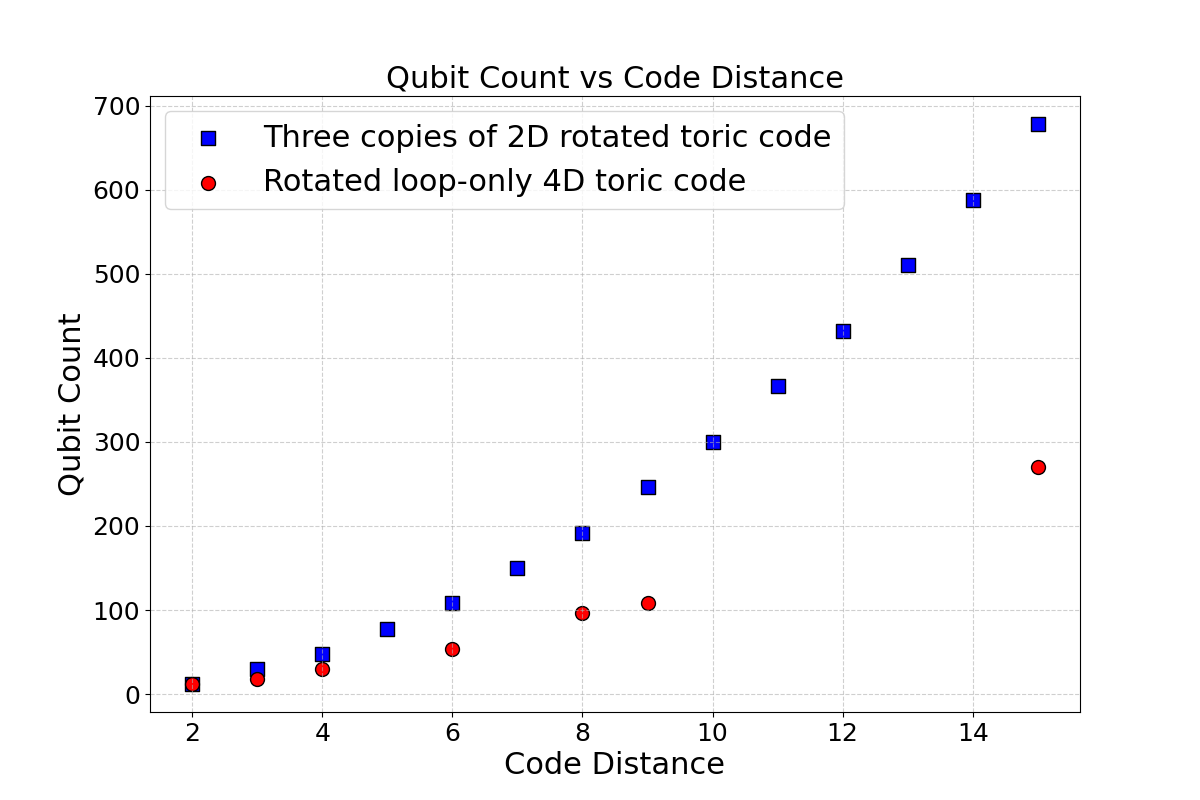}
    \caption{Comparison of qubit counts for rotated 2D and optimally rotated 4D toric codes. 
    For $d$ even we use three copies of 2D toric code with parameters $[[3d^2,6,d]]$ and for $d$ odd we use $[[3d^2+3,6,d]]$. 
    }
    \label{fig:toric-code-figure}
\end{figure}

A natural first point of comparison is the two-dimensional toric code. Without a rotation, the four-dimensional loop-only toric code and two-dimensional toric code have the same encoding rate $[[6d^2,6,d]]$ versus three copies of two-dimensional toric code $ 3\times [[2d^2,2,d]] = [[6d^2,6,d]]$. 
Of course, the loop-only toric code is single-shot, while the two-dimensional toric code is not.
However, once we introduce a rotation, we have a target to beat, if you use rotated two-dimensional toric code you get $3\times[[d^2,2,d]] = [[3d^2,6,d]]$.
In \Cref{fig:toric-code-figure}, we show the qubit count for an optimal rotation versus code distance of the loop-only toric code and a two-dimensional toric code plotted as a baseline. 

Compared to general LDPC codes that are not toric codes, the 4D loop-only toric codes has several desirable features.  The 4D loop-only toric codes on hypercubic lattices are self-dual (up to a permutation of the qubits), so a logical Hadamard (and permutation of the logical qubits) can be implemented by applying a Hadamard to every physical qubit and permuting the physical qubits.  Also the lattice surgery can enable joint measurements of logical operators on two different toric code patches. Finally as mentioned already, the loop-only toric codes are self-correcting memories, and have single-shot properties.

We give four sample codes for performance comparison: 
the Det16 geometric code [[96,6,8]], the rotated surface code [[64,1,8]] from \cite{o2024compare,fowler2009high}, the BB code [[90,8,10]] from \cite{bravyi2024high}, and  the SHYPS code [[225,16,8]] from \cite{malcolm2025computing}.  
Their logical error rates at physical error rate=$10^{-3}$ and pseudo-thresholds are listed in Table ~\ref{table: comparison}.

\begin{table}
\centering
\small
\begin{tabular}{lcccl}
\toprule
Codes for comparison 
&  Code parameters& $P_L(0.001)/k$& $P_L(0.001)$& Pseudo-threshold \\
\midrule
4D geometric code & [[96,6,8]]&   $ 4\times 10^{-7}$ &$2 \times 10^{-6}$& $ 0.01$ \\
Rotated surface code \cite{o2024compare,fowler2009high} & [[64,1,8]]& $ 7\times 10^{-6}$ &$7\times 10^{-6}$& $ 0.0024$ \\
BB code \cite{bravyi2024high} & [[90,8,10]]&  $ 6\times 10^{-7}$ &$5 \times 10^{-6}$& $ 0.0053$ \\
SHYPS \cite{malcolm2025computing} & [[225,16,8]]&  $ 3\times 10^{-5}$ &$5 \times 10^{-4}$& $ 0.003$\\
\bottomrule
\end{tabular}
\caption{Logical error rates per round of error correction at physical error rate=$10^{-3}$ and pseudo-thresholds.
Here, $P_L(p)$ denotes the block logical error rate at physical error rate p; the logical error rate per logical qubit is $P_L(p)/k$, where k is the number of logical qubits. The pseudo-threshold is defined by the condition $P_L(p) = k p$.}
    \label{table: comparison}
\end{table}

\section{Clifford completeness and synthesis}\label{sec: clifford completness}

\subsection{Full set of Clifford logical primitives}

To go beyond logical memory, explicit unitaries for the logical gates are needed in order to obtain a complete generating set for the logical Clifford groups.  For a performance comparison of various codes, it is necessary to have a cost estimate of the logical cycles.  In this section, we show that all 6 logical qubits in a single block of the 4D geometric codes are available for full logical information processing.  

The completeness of logical Clifford operations is mainly a performance issue.  One way to make the code Clifford complete is to perform injection of each operation.  Without any cost constraint, such injections could make all our codes logical Clifford complete.  However, such an approach requires high depth circuits, with lots of physical operations.  Such operations would incur both a performance hit in terms of logical error rate, but also incur a long logical operation time, slowing down overall runtime of the machine.  Instead, we want to identify a set of logical operations that are both efficient in space and time, and also retain adequate logical error rate performance.

Since space group and ZX duality symmetry logical operations are not known to form a complete logical gate set, we will employ lattice surgery to find additional gates.  Our general strategy below completes the standard 4D lattices and the Hadamard lattice.  However, an additional approach is to implement topological gates with low-depth circuits and include them as part of the Clifford synthesis in \Cref{sec: topologicalGates}.  

Our general strategy to achieve Clifford completeness of the single block of a given 4D geometric code consists of three steps.  First we run a best-effort search of the symmetries of the given code block, which can be interpreted as logical operations (we have not yet found any lattice where those symmetries give a Clifford-complete set of logical operations for a single block).  In the second step, we use lattice surgery and partial transversal CNOTs to build targeted CNOTs within a single block with one or two ancillary blocks.  Finally, we combine the resulting logical gates from the previous two steps, and test if they are sufficient for Clifford completeness.  The strategy works well for the standard lattices with unit cells 4-cube $l\times l\times l\times l$.  We also illustrate this strategy with the Hadamard lattice code for rotated lattices below, and prove the Clifford completeness in this case.  We expect that the same strategy works for many more lattices if not all, but leave it as an problem to the future.

\subsubsection{Block layout}

We envision a quantum computer made of multi-blocks of some Det$L$ codes laid out in a plane.  The hardware has all-to-all connectivity and is capable of permuting qubits.  Platforms such as neutral atoms, ion traps, and photonic machines all have such capabilities.

Each block of a 4D geometric code has 6 logical qubits.  For our general theoretical discussion, we assume that the blocks are sitting at the vertices of an $l\times l$ square in the square lattice $\bZ^2$ of the plane $\bR^2$.  

\subsubsection{Permutation and fold-transversal logical operations}

Symmetries of the loop-only Hamiltonian leads to interesting logical operations.  Qubit permutation, such as those coming from lattice-preserving rotations and reflections, can produce non-trivial logical operations that become encoded as CNOTs and SWAPs (logical qubit permutations), but in general it is exponentially hard to identify all such symmetries~\cite{Leon82}. 
 
One interesting construction of symmetries of the loop-only toric code is via the $ZX$ duality introduced in \cite{breuckmann2024fold}.  
A $ZX$ duality of a CSS code is a permutation of the qubits such that the support of the $X$ and $Z$ stablizers are interchanged. While the permutation of qubits in a ZX duality itself is not necessarily a symmetry of the Hamiltonian, two types of symmetries, called Hadamard-type and Phase-type automorphisms can be constructed \cite{breuckmann2024fold}.
An important class of ZX dualities are found by using the four dimensional space group of the crystal upon which the loop-only toric code is constructed. 
All the intrablock symmetries we considered here can be labeled by a four dimensional space group transformation $(M,b)$, where $M$ is an orthogonal matrix, and $b$ is a translation along with whether it is Hadamard-type or Phase-type. 

Given a ZX duality $\tau$, the Hadamard-type symmetry is obtained as 
\begin{align}
H_\tau = \tau \circ \bigotimes_{i} H_i 
\end{align}
where $H_i$ is the Hadamard gate on qubit $i$, and $\tau$ is the qubit permutation. 

If the ZX duality is further of order 2, then the Phase-type symmetry can be obtained:
\begin{align}
S_\tau = \bigotimes_{i = \tau(i)} S_i \bigotimes_{i \neq \tau(i)} CZ_{i, \tau(i)}
\end{align}

\subsubsection{Lattice surgery}

To augment the logical gates from symmetries, we use lattice surgery. Lattice surgery can prepare the necessary resource states to implement the partial transversal CNOTs from \cite{goto2024high}, which in turn leads to new targeted intrablock CNOTs, and Clifford completeness of a single block.

Our surgery is a generalization of the lattice surgery in 2D to 4D and multiple logical qubits.  The effect of a lattice surgery is more complicated and the detail is in \cite{aasen2025topological}.
 
Using lattice surgery, we can measure three logical qubits of a code block in a given basis to prepare logical states of the form $|+++000\rangle$.
Then using such states and the partial transversal CNOTs in \cite{goto2024high}, we can perform a CNOT between three qubits of one code block and three qubits of a second one, with all those three qubits having one direction in common.  By using those CNOTs repeatedly, with different common directions, we may perform a targeted CNOT between a single qubit in one code block and the same one in a different block. More detail is given below.  

\subsubsection{Targeted CNOT}

For general 4D lattices, explicit bases of the logical qubits are nontrivial to find, and the detail is in \Cref{logical basis}.  Below we describe the second step using the standard lattices.  The 6 logical qubits are indexed as follows: every two directions $xy,x,y\in \{0,1,2,3\}$ of the 4 coordinates represent a torus support of a logical operator.  For specificity, we fix them as 6 independent logical logical operators, and hence a basis of the 6 logical qubits.  The first three are of the form $0a$ for $a \in\{1,2,3\}$ and the second three are of the form $bc$ for $b,c\in \{1,2,3\}$.   

The second step of our general strategy has three parts.

In part iia), we use surgery to create 4 pairs of resource states by cutting the 4-torus along one of the 4-directions $\{0,1,2,3\}$.  In general, any cut would work as well except then no obviously well-behaved logical bases under surgery.  In more detail, 
we first use surgery to cut open the 4-torus $T^4$ into a slab $T^3\times [0,1]$ of the 3-torus along one of the coordinate directions.  With appropriate boundary conditions of either Z or X on the boundary 3-tori, this operation allows us to prepare fault-tolerantly either the logical $\ket{+++000}$ or $\ket{000+++}$ states as desired.  

In part iib), we make use of the partial transversal CNOTs in \cite{goto2024high} and the desired states created earlier to produce partial transversal CNOTs of logical qubits of the form $0a$ from one block to another.  Note that the partial transversal CNOT in \cite{goto2024high} requires a Hadamard, and the Hadamard lattice code has only a Hadamard combined with a SWAP.  However, this is not an issue since the Hadamard gate is needed only immediately after preparation or immediately before measurement.  There are multiple ways to deal with this issue. One way is to simply account the extra SWAPs in the decoder.  Another is to prepare either $\ket{+++000}$ or $\ket{000+++}$, the Hadamard could be replaced by a change in choice of preparation state if it is immediately after preparation, or by a change in choice of measurement basis if it is immediately before measurement.

In part iic), we take 3 blocks of the code, called them 1,2,3.  Block 2 is prepared in the $\ket{000000}$ state. We do partial transversal CNOT of qubits of form $0a$ from block 1 to block 2,  then do partial transversal of qubits of form $1b$ from block 2 to block 3, followed with partial transversal of qubits $0a$ from block 1 to block 2.  This protocol results in a partial transversal CNOT of $01$ from block 1 to block 3.

Finally for the third step of our general strategy, we build new gates for a single block using the resulting targeted CNOTs from Step 2 and the symmetry logical gates from Step 1 to obtain the full logical Clifford operations.

\subsubsection{Clifford completeness}

All three general steps for Clifford completeness depend on many choices such as bases of logical qubits and directions to cut for surgery, hence general conclusions independent of lattices are challenging to prove. Below we carry out the general strategy with some variations specific to the Hadamard lattice and expect a similar proof for at least all odd determinant lattices.  

A single block of the Hadamard lattice code is a [[96,6,8]] code with $64$ X-stabilizers and $64$ Z-stabilizers.  The Hadamard lattice is among the most symmetric with 384 lattice automorphisms, which yield 24 distinct permutation symmetries.  The logical gates from permutation symmetries and fold-transversals form a subgroup $G_H$ of the full 6-qubit Clifford group of order=$2^7\times 3^2 \times 5\times 7\times 13\times 19\times 31$, which is only a small fraction of the full 6-qubit Clifford group  \cite{aasen2025topological},  but essentially any additional gate would complete $G_H$ to the full 6-qubit Clifford group.

For the Hadamard lattice codes, a new intrablock logical CNOT gate can be constructed after Steps 1 and 2 to complete the full logical Clifford operations of the six logical qubits in a single block.

We leave out the full detail as the Clifford synthesis below gives a different algorithmic proof of the Clifford completeness for the Hadamard code.

\subsection{Clifford circuit synthesis}

We provide algorithms to find circuits for
\begin{itemize}
    \item multi-qubit Pauli measurements, 
    \item multi-qubit Pauli exponents of the form $\exp(i\phi P)$, 
    \item multi-target CNOTs, and 
    \item common one- and two-qubit Clifford unitaries 
\end{itemize}
in terms of primitive operations of CSS block codes.
We focus on the geometric code with Hadamard lattice, \cref{eq:hadamard-lattice-hnf}, which encodes six qubits per block, 
but the techniques we describe here apply to other lattices and codes.
Our algorithm for synthesis of multi-qubit Pauli measurements and Pauli exponents can readily be used to
compile any quantum algorithm via Pauli-based computation~\cite{Bravyi2016} into sequences of primitive fault-tolerant operations.
Additionally, multi-qubit Pauli measurements can be used to implement distillation protocols~\cite{Haah2018codesprotocols,Litinski2019gameofsurfacecodes,Litinski2019magicstate}, and multi-target CNOTs are commonly used in table-lookup circuits which in turn are common in quantum algorithms for chemistry~\cite{haner2022spacetimeoptimizedtablelookup}.

The new techniques are needed for two reasons. First, primitive logical unitaries within one block of the codes we consider here do not generate the full Clifford group. Second, the only measurements available via surgery measure a group of commuting Pauli operators with six generators. Our techniques provide flexible space-time trade-offs. Many of the choices can be refined based on a more detailed analysis of performance of logical operation. 

We also provide circuit costs for familiar one- and two-qubit Clifford unitaries within and across blocks 
as an evidence of universality of primitive operations, but we emphasize that more general Clifford operations on 6 or 12 qubits 
may be synthesized directly using much fewer logical primitives. Generally the number of error-correction cycles $N_{\mathrm{EC}}$ needed for common one and two-qubit gate is not a representative metric of the synthesis overhead for codes that encode $k$ logical qubits per block. This is because implementing one and two-qubit operations requires space-time volume $k N_{\mathrm{EC}}$ thus penalizing error-correcting codes with larger $k$. For this reason, we mainly focus on higher-level primitives spanning multiple blocks.

\subsubsection{Multi-qubit Pauli measurements}

We synthesize a multi-qubit measurement $M_P$ of Pauli operator $P$ in two steps.
We first find a circuit for multi-qubit tensor product of logical $Z$ operators acting on the first logical 
qubit of each consecutive pair of block in a sequence of $2N$ blocks (i.e., the operator $\mathcal{Z}=\bigotimes_{i=1}^{N} Z_{12i-11}$).
Second we find Clifford unitaries $C_j$ for each consecutive pair of blocks in the support that map $Z_1$ to 
the $P_j$, the restriction of $P$ to qubits in the blocks $2j-1,2j$.
The measurement $M_P$ is then implemented by applying $C_j^\dagger$ on each pair of blocks, followed by $M_{\mathcal{Z}}$, 
followed by $C_j$ on each block.

\begin{figure}
    \centering
    \includegraphics[width=0.85\linewidth]{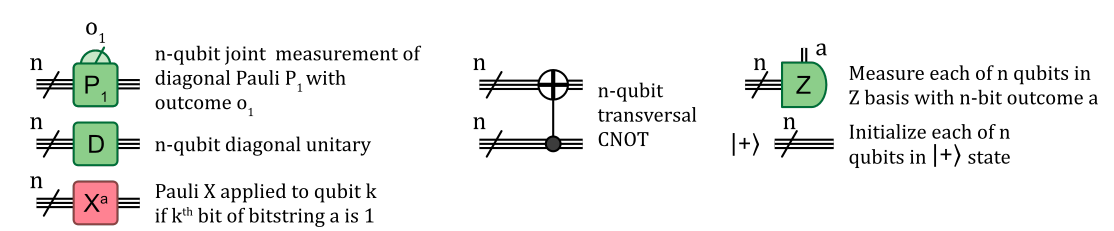}
    \caption{Notation for transversal gates.
 }
 \label{fig:transversal-circuit-notation}
\end{figure}

\begin{figure}
    \centering
    \includegraphics[width=0.75\linewidth]{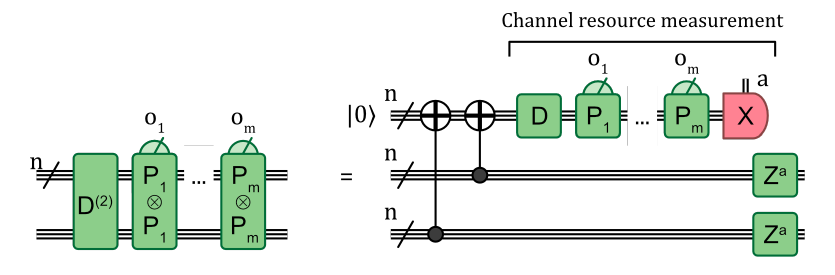}
    \caption{Multi-target diagonal channel ejection useful for implementing channels symmetric across several blocks. Unitary $D$ is diagonal in $Z$ basis, $P_1,\ldots,P_m$ are multi-qubit $Z$ measurements.
    Unitary $D^{(2)}$ is given in \cref{eq:block-symmetric}. }
    \label{fig:multi-target-ejection}
\end{figure}

\begin{figure}
\centering
\begin{subfigure}[b]{0.45\textwidth}
\includegraphics[width=\linewidth]{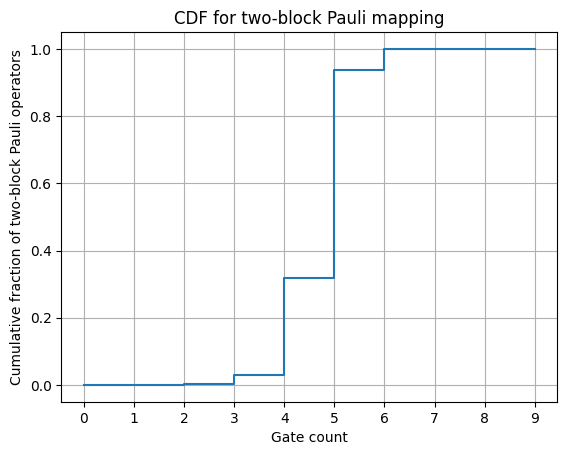}
\caption{}
\end{subfigure}~
\begin{subfigure}[b]{0.45\textwidth}
\includegraphics[width=\linewidth]{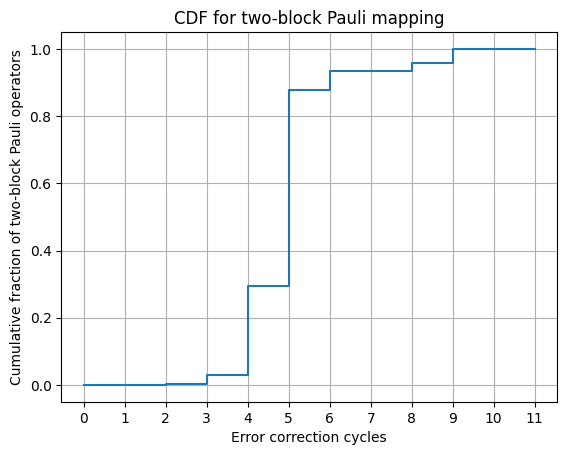}
\caption{}
\end{subfigure}
    \caption{The number of (a) logical gates and (b) error correction cycles to map $Z_1$ to an arbitrary Pauli operator $P$ on twelve qubits. The vast majority of of Pauli operators ($\sim93\%$) 
    do not require a gate that uses diagonal injection/ejection, while the remaining $\sim7\%$ of Pauli operators require only one such gate.
    The number of error correction cycles is larger than the gate count because it accounts for the cycles needed to prepare a resource state on ancillary qubits.}
    \label{fig:pauli-mapping-cost-distribution}
\end{figure}

Using the diagramatic convention of \Cref{fig:transversal-circuit-notation}, the projective measurement $M_{\mathcal{Z}}$ can be implemented with an ancilla block as illustrated in \Cref{fig:multi-target-ejection}. 
We formalize the task of finding the unitaries $C_j$ as a Pauli mapping problem: given Pauli unitaries $P$ and $Q$, 
find a circuit consisting of primitive operations that implements unitary $C$ such that $CPC^\dagger = Q$.
Unitary operations acting on one block do not generate the six-qubit Clifford group in case of the Hadamard lattice, and therefore we cannot solve the Pauli mapping problem by relying on unitary logical primitives acting on one block. We checked this using a simple breadth-first search, with at most $4^{6}=4096$ nodes being Pauli operators and edges being one-block primitive unitary operations (a similar statement can be made about unitary logical primitives acting on two blocks).
We get around this limitation by using a diagonal channel injection/ejection technique that relies on an additional ancilla block, resource states, and transversal CNOTs. 
The ejection circuits for applying multi-qubit diagonal unitaries were introduced in Ref.~\cite{EDP2022} under the name diagonal remote execution gadgets.
The same techniques are used to to implement logical multi-qubit Z measurements. The resulting cost distribution of solutions to the Pauli mapping problem is given in \Cref{fig:pauli-mapping-cost-distribution}.

\subsubsection{Multi-qubit Pauli exponents}
\label{sec:multi-qubit-pauli-exponentials}

We implement multi-qubit Pauli exponents in two steps, in a similar way to how we implement multi-qubit Pauli measurements. We rely on diagonal unitary ejection from ~\Cref{fig:diagonal-channel-ejection} to implement $\exp(i\phi \mathcal{Z})$ for $\mathcal{Z}=\bigotimes_{i=1}^{N} Z_{12i-11}$, which applies to non-Clifford diagonal unitaries. Then we transform $\mathcal{Z}$ to required Pauli operator via Pauli mapping. 

\paragraph{Batch Pauli mapping}
When implementing several Pauli exponents consecutively, we can reduce the cost of Pauli mapping. Suppose we would like to implement product 
\begin{align}
\exp(i \phi_1 P_1) \ldots \exp(i \phi_m P_m) 
\end{align}
via a choice of Clifford unitaries $C_1, \ldots, C_{m+1}$ 
and fixed Pauli exponents $\exp(i \phi_1 Z_1)$
\begin{align}
C_1~\exp(i \phi_1 Z_1)~C_2~\exp(i \phi_2 Z_1)~\ldots~C_m~\exp(i \phi_m Z_1)~C_{m+1} 
\end{align}
Unitaries $C_j$ must satisfy the following equations 
\begin{align}
C_1 Z_1 C_1^\dagger &= P_1,\\
C_2 Z_1 C_2^\dagger &= C_1^\dagger P_2 C_1,\\
C_3 Z_1 C_3^\dagger &= (C_1 C_2)^\dagger P_3 (C_1 C_2),\\ 
&\vdots\nonumber
\end{align}
along with
\begin{align}
C_1 C_2 \cdots C_m &= C_{m+1}^\dagger.
\end{align}
The above idea applies also applies in the setting we use for multi-qubit Pauli 
measurements and Pauli exponents, with Clifford unitaries chosen per block of $12$ qubits.
In this case, each of unitaries $C_j$ can be found using a breadth-first search
and follow distribution in~\Cref{fig:pauli-mapping-cost-distribution}. Without this batching approach the overall cost of the sequence of Pauli exponents would be roughly doubled since each Pauli mapping would have to be implemented and then reversed before the next Pauli exponent can be applied.

\begin{figure}
    \centering
    (a) \includegraphics[width=0.45\linewidth]{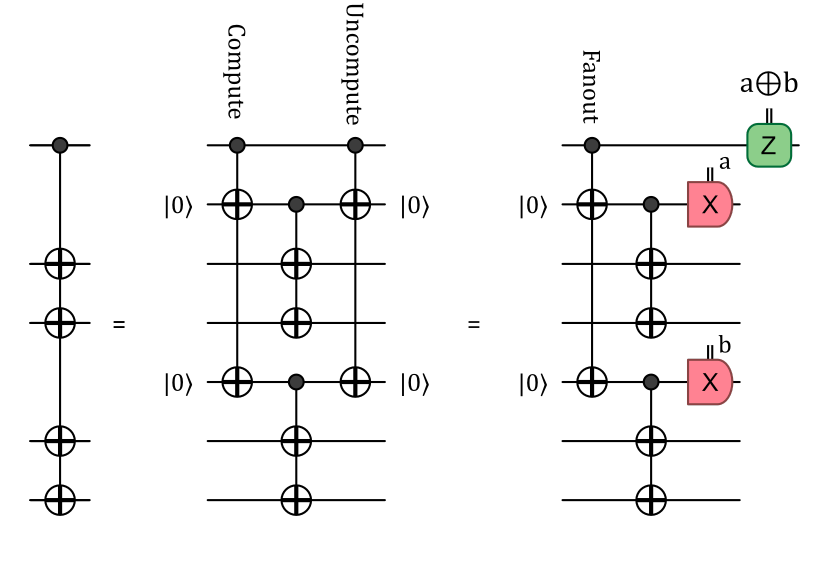}
    (b) \includegraphics[width=0.45\linewidth]{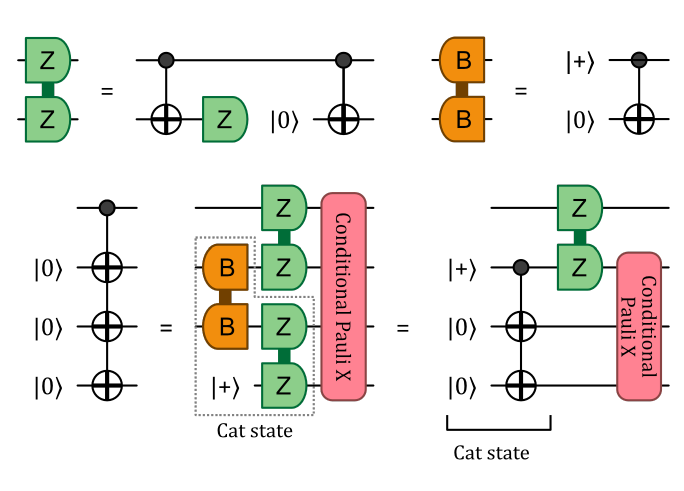}
    \caption{(a) Multi-target CNOTs can be implemented using fan-out gate and single qubit measurements. (b) Fan-out gate can be implemented using Bell state preparations and non-destructive ZZ measurements in constant depth.}
    \label{fig:multi-target-cnot}
\end{figure}

\paragraph{Logical space-time volume of a Pauli exponent}
We provide an example of a logical space-time volume estimate for a Pauli exponent implemented via multi-target diagonal channel ejection~(\Cref{fig:multi-target-ejection}) to illustrate available space-time trade-offs. We use two ancillary blocks per two target blocks of the Pauli exponent. One block is used for constant-depth multi-target transversal CNOT~(\Cref{fig:multi-target-cnot}) and another one is 
used for Pauli mapping. Assume that transversal CNOT, transversal state preparation in $|0\rangle$, $|+\rangle$ and Bell state take one logical cycle and transversal destructive measurement take $0$ logical cycles. Constant depth fan-out gate requires three cycles, and multi-target CNOT with one ancilla per two targets requires five cycles. Pauli mapping requires at most nine logical cycles as illustrated in \Cref{fig:pauli-mapping-cost-distribution}(a), taking into account that resource state for a diagonal gate used during mapping is prepared in parallel with execution of fan-out gate, and using batch Pauli mapping described above. 
In summary, space-time volume per Pauli exponent target is $28$.

\subsubsection{Diagonal channel injection and ejection}
\label{sec:diagonal-channels}

\begin{figure}
    \centering
    \includegraphics[width=0.75\linewidth]{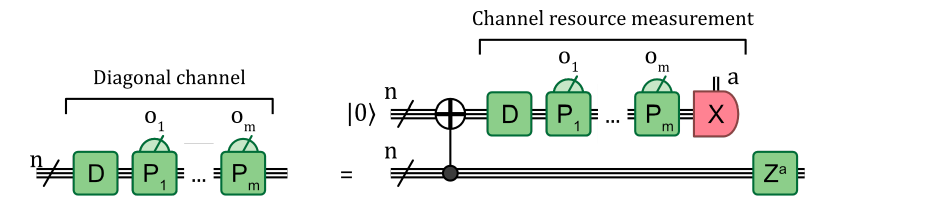}
    \caption{Diagonal channel ejection protocol implements an $n$-qubit diagonal channel using $n$ qubits prepared in $|+\rangle$, 
transversal CNOT and a resource measurement followed by Pauli corrections.
Unitary $D$ is diagonal, $P_1,\ldots,P_m$ are multi-qubit diagonal Pauli measurements.  See~\cref{fig:transversal-circuit-notation} for circuit diagram notation.}
    \label{fig:diagonal-channel-ejection}
\end{figure}

\begin{figure}
    \centering
    \includegraphics[width=0.75\linewidth]{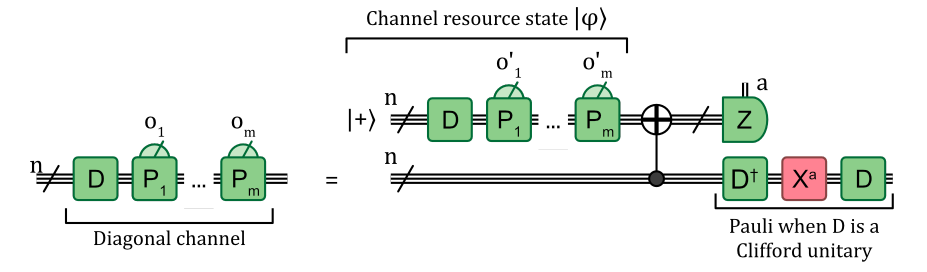}
    \caption{Diagonal channel injection protocol implements a diagonal channel using a resource state $D M_{P_1} \ldots M_{P_m} | + \rangle ^{\otimes n}$, transversal CNOT, 
transversal $Z$ measurement
and corrections $X^a D X^a D^\dagger$, where $X^{a} = \prod X_j^{a_j}$ for $a \in \{0,1\}^n$.
Unitary $D$ is diagonal, $P_1,\ldots,P_m$ are multi-qubit diagonal Pauli measurements. Channel outcomes 
$o_j$ and $o'_j$, related via $o'_j = o_j + [P_j, X^a]$ where $[P,Q]$ is zero when Paulis $P,Q$ commute and is one otherwise. See~\cref{fig:transversal-circuit-notation} for circuit diagram notation.
 }
 \label{fig:diagonal-channel-injection}
\end{figure}

A channel is $Z$-diagonal if it consist of a unitary $D$ diagonal in $Z$ basis followed by a sequence of multi-qubit $Z$-diagonal projective measurements $P_1,\ldots, P_m$.
$X$-diagonal channels are defined similarly. 
We provide circuits that implement diagonal $Z$ channels via resource measurements (\Cref{fig:diagonal-channel-ejection}), resource states (\Cref{fig:diagonal-channel-injection}), transversal CNOTs, transversal $X/Z$ state preparations, and measurements commonly available in CSS codes encoding multiple logical qubits. 
Similar circuits are available for $X$-diagonal channels.

The resource state for channel $D M_{P_1} \ldots M_{P_m}$ can be used to implement channel 
$$
D^{(N)} M_{P^{\otimes N}_1} \ldots M_{P^{\otimes N}_m}
$$  on $N$ target blocks, where $D^{(N)}$
and $D$ are related as following:
\begin{equation}
\label{eq:block-symmetric}
      D =  \sum_{k \in \{0,1\}^n} e^{i \phi_k } |k\rangle \langle k|, \, D^{(N)} = \sum_{k \in \{0,1\}^n} e^{i \phi_k } (|k\rangle \langle k|)^{\otimes N},
\end{equation}
as shown in \Cref{fig:multi-target-ejection}.

We use multi-target ejection to implement a multi-block joint Pauli $Z$ measurement via resource state for measuring $Z_1$ on one block. Similarly multi-target ejection can be used to implement generalized $S$ gate 
$\exp\!\left(-i \frac{\pi}{4} Z^{\otimes 6 N}\right)$ by using resource state for  $\exp\!\left(-i\frac{\pi}{4} Z^{\otimes 6}\right)$. Moreover, we can similarly implement $X$ or $Z$ diagonal Pauli exponents. 

Multi-target injection is also available, similar to multi-target ejection. 
The sequence of transversal CNOTs used in multi-target $X$-diagonal injection/ejection protocol are transversal multi-target CNOTs. 
They can be implemented using ancillary qubits in constant depth. 
We discuss various space-time trade-offs for multi-target CNOT implementation in the next section.
In the $Z$-diagonal case the sequence of CNOTs implements a CNOT with a single
target controlled on parity of several qubits, we call these gates parity-controlled NOTs.

\subsubsection{Multi-target CNOTs and parity-controlled NOTs}
\label{sec:multi-target-cnots}

In this section we consider both transversal and selective multi-target CNOTs. 
We focus on multi-target CNOTs, but all the results also apply to parity-controlled NOTs. This is because
parity-controlled NOT is the product $\prod_{j=1}^{N} \mathrm{CX}_{j,N+1}$
and multi-target CNOT is the product $\prod_{j=1}^{N} \mathrm{CX}_{N+1,j}$.
Both gate are related via conjugation by a tensor product of Hadamard gates because 
$(H \otimes H) \mathrm{CX}_{i,j} (H \otimes H) = \mathrm{CX}_{j,i}$.

There are three options for implementing multi-target CNOT. Two purely unitary options are sequential (linear depth) and tree-like (logarithmic depth). The third option uses one-qubit $X$, $Z$ measurements and ancillary qubits to implement CNOT in constant depth~(\Cref{fig:multi-target-cnot}). Moreover, we can interpolate between the three options for space-time trade-offs.

\begin{figure}
    \centering
    (a) \includegraphics[width=0.45\linewidth]{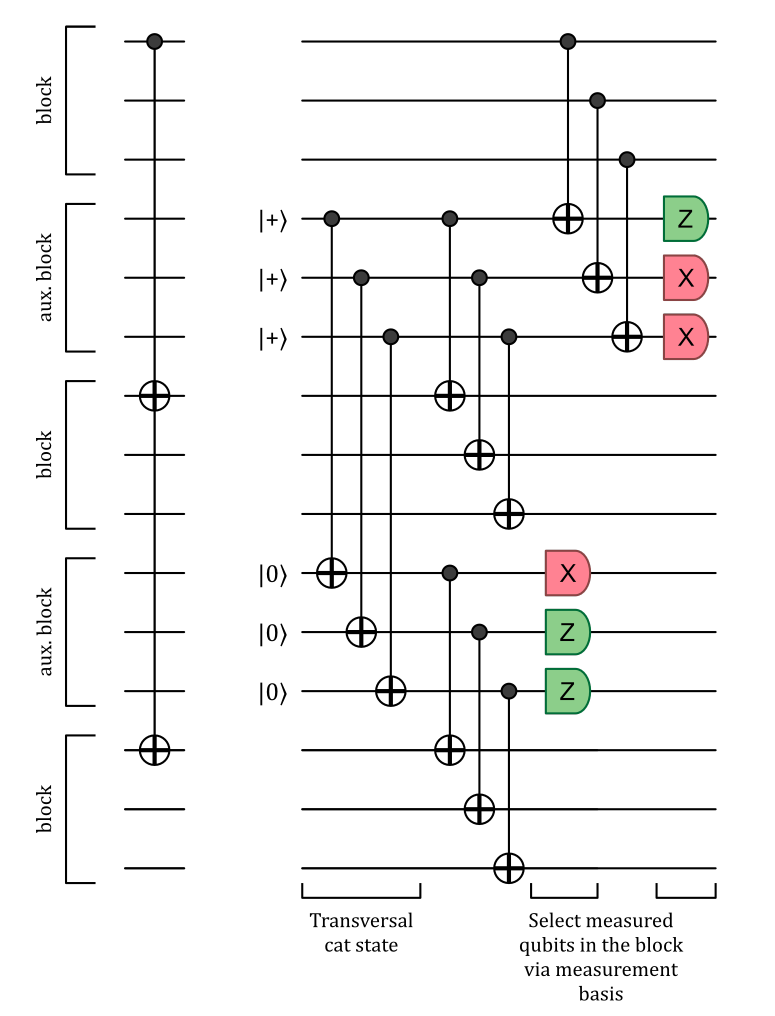}
    (b) \includegraphics[width=0.45\linewidth]{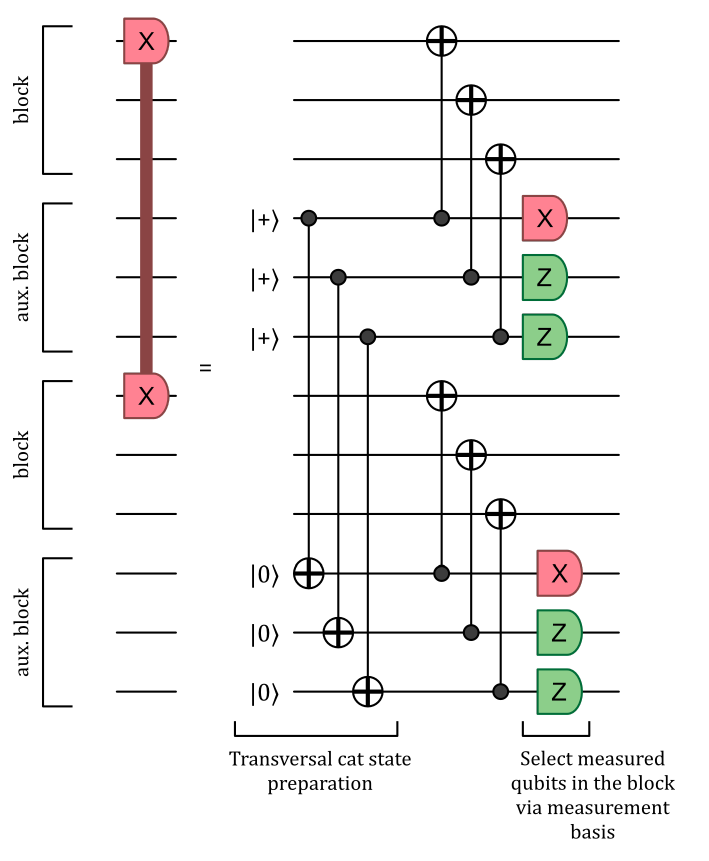}
    \caption{(a) Constant-depth cross-block multi-target CNOT and (b) cross-block multi-qubit $X$ measurement. }
    \label{fig:cross-block-multi-target-cnot}
\end{figure}

For injection/ejection circuit transversal multi-target CNOTs are sufficient. 
For algorithms like table-lookup we need cross-block multi-target CNOTs. They can be implemented in constant depth by preparing transversal cat-state and selectively measuring qubits in $X$ or $Z$ bases~(\Cref{fig:cross-block-multi-target-cnot}). A single target within a block can be transformed to multiple targets via Pauli mapping, similar to how we implement Pauli exponents and Pauli measurements. When implementing several multi-target CNOTs, we can take advantage of Batch Pauli Mapping discussed in \Cref{sec:multi-qubit-pauli-exponentials}.

\subsubsection{Synthesis of resource states}

We expand resource states that can be prepared with primitive operations using lattice surgery. 
There are four additional states that one can prepare starting with two blocks initialized in a plus state, 
applying surgery and destructively measuring X on all qubits in one of the blocks. 
We assume that these state preparations can be further simplified to require 
three EC cycles and no additional blocks.
These four states followed by fold-$S$, fold-$S_x$, and a qubit permutation gates give us access to resource states 
for diagonal unitaries that extend one-block primitive operations to a set that generates the full Clifford group.
We consider all resource states that use three fold-$S$, fold-$S_x$
gates. There are $2112$ such resource states that correspond to diagonal gates in $X$ or $Z$ basis. 

To enumerate available resource states we implement standard breadth-first search for stabilizer state preparation.
Each node is $n$-qubit stabilizer state, represented as $n\times 2n$ binary matrix in row-reduced echelon form.
The total number of stabilizer states on six qubits is relatively small and such a search can be run on a laptop.

\paragraph{Measurement resource states}

For measurements of Pauli operators diagonal in $X$ or $Z$ basis we define the corresponding resource state using the diagonal channel injection circuit in~\Cref{fig:diagonal-channel-injection}. 
For example, for a measurement diagonal in $Z$ basis it is a resource state obtained by applying the measurement to a register with qubits in a plus state.
Preparing resource states for measurements requires one diagonal gate that itself uses a resource states obtained using surgery and fold gates.
In \Cref{tab:measurement-resource-states} we summarize the costs and structure of the circuits for preparing various resource states.

\begin{table}
    \centering
\begin{minipage}{0.45\textwidth}
\centering
\begin{tabular}{lcc}
\toprule
Pauli &  Gate count & $\Delta$CX \\
\midrule
$Z_0$ & 8 & 3 \\
$Z_1$ & 5 & 4 \\
$Z_2$ & 9 & 5 \\
$Z_3$ & 10 & 5 \\
$Z_4$ & 8 & 3 \\
$Z_5$ & 5 & 4 \\
$Z_0 Z_1 Z_2 Z_3 Z_4 Z_5$ & 6 & 2 \\
\bottomrule
\end{tabular}
\end{minipage}
\begin{minipage}{0.45\textwidth}
\centering
\begin{tabular}{lcc}
\toprule
Pauli &  Gate count & $\Delta$CX \\
\midrule
$X_0$ & 5 & 4 \\
$X_1$ & 8 & 3 \\
$X_2$ & 10 & 5 \\
$X_3$ & 9 & 5 \\
$X_4$ & 5 & 4 \\
$X_5$ & 8 & 3 \\
$X_0 X_1 X_2 X_3 X_4 X_5$  & 6 & 2 \\
\bottomrule
\end{tabular}
\end{minipage}

    \caption{Costs of resource states for targeted non-destructive measurements. Each resource state requires one diagonal gate applied using 
    a resource state. Gate count is a total number of gates needed, $\Delta$CX 
    is the number of gates applied before transversal CNOT is used to apply a diagonal gate from a resource state.}
    \label{tab:measurement-resource-states}
\end{table}

\subsubsection{Circuits for common one- and two-qubit Clifford unitaries}

Here we summarize the cost of implementing common one- and two-qubit Clifford unitaries to illustrate the Clifford-completeness of the logical primitives described here. As discussed before, the costs described here are not representative of the synthesis overhead in an algorithm.

\textbf{Targeted $S$ ($\sqrt{Z}$) gates.} There are two logical $S$ gates that are products of four fold-$S$ gates available on every block of six qubits -- these are the logical  $S_2$ and $S_6$. 
All other $S$ gates require auxiliary block to be implemented.
Preparing resource states for four other $S$ gates requires $6$ fold 
gates and one diagonal gate obtained from surgery.
Diagonal gate obtained from surgery use resource state or resource measurement that requires $6$ EC cycles, therefore each $S$ gate except $S_2$, $S_6$ requires $9$ EC cycles and two auxiliary blocks.

\textbf{Targeted $S_x$ ($\sqrt{X}$) gates.}
There are two logical $S_x$ gates that are products of four fold-$S_x$ gates available on every block of six qubits -- these are the logical $S_{x,1}$ and $S_{x,5}$ gates. All other $S_x$ gates are implemented via resource states 
and require $9$ EC cycles and two auxiliary blocks,
similar to $S$ gates.

\textbf{Targeted $S_y$ ($\sqrt{Y}$) gates.}
We obtain all targeted $S_y$ gate by conjugating 
$S_1$ with a Clifford that maps $Z_1$ to $Y_j$.
Mapping $Y_1, Y_2, Y_5, Y_6$ requires 
four fold gates, the rest require one diagonal gates obtained from surgery and three fold $S$ gates.
The targeted $S_y$ gates require $17$ or $25$ EC cycles and two auxiliary blocks.

\textbf{Targeted in-block CZ gates} There is a logical $CZ$ gate on qubits $2$, $6$ that can be implemented by a product of $6$ fold-$S$ gates. All other $CZ$ gates require two auxiliary blocks to be implemented. Resource states for $CZ$ gates use one or two diagonal gates obtained from surgery and require $10$ to $16$ EC cycles and one auxiliary block. An additional auxiliary block is needed to implement the gate via a resource state.

\textbf{Targeted in-block CNOT gates} 
We obtain all targeted in-block CNOT gates 
with control $i$ and target $j$
by conjugating CZ on qubits 1,2 by a Clifford that maps group generated by $Z_1,Z_2$ to the group generated by $Z_i,X_j$. 
CZ on qubits 1,2 requires $10$ EC cycles. 
The conjugating Clifford unitaries require between 8 and 21 EC cycles and overall targeted CNOTs within a block require between $26$ to $52$ EC cycles.

\textbf{Targeted inter-block CNOT gates} 
Targeted inter-block CNOTs are a special case of multi-target CNOTs discussed in
\Cref{sec:multi-target-cnots}.

\subsection{Universality}

To achieve fault-tolerant universality, we will employ the standard state injection and magic state distillation techniques \cite{knill2004fault,bravyi2005universal}.  A generalization of state injection to the case for multiple qubits is given in Ref.~\cite{aasen2025topological}.

\section{Realizing a fault-tolerant quantum computer with 4D geometric codes}

Universal fault-tolerant quantum computers may be realized using 4D geometric codes,
which are designed to enable efficiently realizing an increasing number of logical qubits with a modest number of physical qubits, while enabling low-depth logical cycles and universal fault tolerance. 
They are amenable to quantum hardware with low-cost all-to-all connectivity and qubit permutation; trapped ions, neutral atoms, and photonic architectures are among such platforms.
The so-called 4D surface code---the open boundary version of 4D loop-only toric code---has been implemented on trapped-ions \cite{berthusen2024experiments};
future realization of our 4D geometric codes offers efficient implementation given its 6-fold reduction in physical qubits required per logical qubit due to the rotation of the code.
In Table~\ref{table:specs} we summarize the specifications required to achieve 50, 100, and 1500 logical qubits using a number of 4D geometric code blocks at two different code distances.

\begin{table}
\centering
\begin{tabular}{rllll}
\hline
4D GC specifications & Parameters &Hadamard& $\ted45$ & Utility\\
\hline
\textrm{Code parameters} & [[$6\cdot$Det,6,$d$]] & [[96,6,8]] & [[270,6,15]] & [[270,6,15]]\\
\textrm{Number of code blocks} & $b$ & 9 & 16 & 250\\
\textrm{Number of logical qubits} & $6\cdot b$ & 54 & 96 & 1500 \\
\textrm{Total data qubits} & $6\cdot b\cdot \ted$ & 864 & 4320 &67500  \\
\textrm{Total measurements per syndrome cycle} & $8\cdot b\cdot \ted$ &1152 & 5760 & 90000\\
\textrm{Depth of starfish syndrome extraction} & 16& 16& 16& 16 \\
\textrm{Depth of compact syndrome extraction} & 8 & 8 & 8& 8 \\
\hline
\end{tabular}
    \caption{Specifications for several 4D geometric codes with varying code distances (4D GC).}
    \label{table:specs}
\end{table}

Using the Hadamard code, a 54-logical qubit quantum computer may be realized in  $2000$ physical qubits or fewer.
The Hadamard lattice has determinant 16, resulting in 96 data and 128 ancilla qubits per block. 
To achieve 54 logical qubits requires 9 blocks of the code and in turn $16\times 14\times 9=2016$ physical qubits.
Simulation indicates the logical error rate per logical qubit per round of error correction is around $10^{-6}$ for physical noise around $10^{-3}$, see Fig.~\ref{fig:be-bposd-monolithic}. 
To further reduce the physical qubit count, blocks of ancilla qubits may be reused at the cost of additional timesteps per logical syndrome extraction cycle.

To realize 96 logical qubits, 16 blocks of the $\ted45$ code may be used, 
requiring $45\times 14\times 16=10,080$ physical qubits.
Based on simulation, the logical error rate per logical qubit per round of error correction will be around $10^{-10}$ for a physical error rate of around $10^{-3}$, see Fig.~\ref{fig:be-bposd-monolithic}.

For realizing upwards of $1000$ logical qubits at utility scale \cite{gidney2025factor,zhou2025resource}, hardware modules each realizing many blocks of a 4D geometric code may be networked together.
Assuming each module has $\sim 10^{5}$ physical qubits, then a networked architecture of modules to reach $\sim1500$ logical qubits in a $\ted45$ code would require $157500$ physical qubits distributed across $\sim 10$ modules.
Here again, the logical error rate per logical qubit per round of error correction is estimated to be around $10^{-10}$ for a physical error rate of around $10^{-3}$.

To transition from proof-of-principle to utility-scale quantum computing, a series of demonstrations are proposed using 4D geometric codes.
Each demonstration incrementally increases the system’s complexity while validating essential components of a fault-tolerant quantum computer.
As a first demonstration, we propose single-shot logical entanglement, such as logical Bell state preparation shared between a pair of code blocks.
Implementations in $\ted9$ and subsequently $\ted16$ codes would highlight both the single-shot capabilities and code scaling.
Beyond entanglement, executing shallow-depth logical circuits involving both intra-block and inter-block operations will be critical for demonstrating computational functionality beyond state preparation.
A follow-on demonstration is to validate the logical memory of the machine, targeting the preservation of $12$ to $18$ logical qubits in a $\ted16$ code over approximately $100$ rounds of quantum error correction.
Finally, to achieve full universality, we propose demonstrating non-Clifford operations via magic state injection and a small-scale distillation protocol.
This step would complete the core set of ingredients needed for a fully functional fault-tolerant quantum computer using a 4D geometric code.
These foundational experiments set the stage for scaling up to larger systems with deep logical circuits and more complex logical interactions.

\section{Discussion and outlook}

While quantum computers have already been built in a variety of platforms, engineering a fault-tolerant, useful quantum computer that outperforms both noisy intermediate scale quantum computers (NISQ) and classical compute remains a significant challenge.
Overcoming the challenge requires enabling both larger numbers of logical qubits and deeper and more complex logical computation. 
Near-term fault-tolerant quantum computers will need to scale to upwards of 50 logical qubits with the ability to perform thousands of fault-tolerant logical operations.
Utility scale quantum computers will require more than a thousand high fidelity logical qubits and upwards of billions of fault-tolerant logical operations.
Transitioning beyond NISQ requires quantum error correction codes and fault-tolerance as a critical part of the quantum software stack.
Our 4D geometric codes satisfy the desirable properties of a fault-tolerant quantum computer: single shot error correction and decoding, high encoding rate, good performance, high threshold error rate, and universal logical operations.  
Two instances of 4D geometric codes, namely the Hadamard and Det45 codes, are presented in detail herein, and present an efficient path toward realizing universal fault tolerance in practice.

There remain many interesting open questions and future directions toward realizing quantum fault tolerance and scale, including:  
\begin{enumerate}
\item  Does there exist a local low-depth circuit implementation of the topological operations in Sec.~\ref{sec: topologicalGates}? Do they enlarge the symmetry logical gates into Clifford completeness? If so, how much do these new logical gates improve the overhead of Clifford synthesis?
\item Prove Clifford completeness of odd-determinant 4D geometric codes as outlined in Sec.~\ref{sec: clifford completness}.
\item Prove or disprove the optimal lattice conjecture in Sec.~\ref{sec: optimal lattice conjecture}.
\item How much does pre-/post-selection enhance quantitatively the performance of the 4D geometric codes?  
\item How much can the 4D geometric codes be sub-systematized to improve code  performance and reduce Clifford synthesis cost?  For example, the Det16 code has distance 8 as a stabilizer code, but distance 9 as a subsystem code \cite{aasen2025topological}.  
\item Is it possible to obtain full logical Clifford completeness for a large subset of the six logical qubits using only transversal gates, say 3 out of 6?  We know at least one among the six logical qubits in a block has full transversal Clifford logical gates \cite{aasen2025topological}. 
\item Is it possible to further reduce the cost of logical Clifford synthesis and universality for 4D geometric codes, in turn reducing the logical operation depth?
\end{enumerate}



\newpage
\appendix
\section{Syndrome extraction circuits}
\label{app:secircuits}
We provide the starfish ordering in Circuit~\ref{starfish} and the compact syndrome extraction circuit in Circuit~\ref{compactse}.

\begin{circuit}
\caption{\label{starfish}Starfish syndrome extraction circuit}
\begin{algorithmic}[1]
\FOR{each $a$ in 1-cells}
    \STATE $\text{PrepX}(a)$
\ENDFOR
\FOR{each $\text{direction}$ in $\{+0,-0,+1,-1,+2,-2,+3,-3\}$}
    \FOR{each $a$ in 1-cells such that $a+\text{direction} = b$ in 2-cells}
        \STATE $\text{CNOT}(a, b)$
    \ENDFOR
\ENDFOR
\FOR{each $a$ in 1-cells}
    \STATE $\text{MeasX}(a)$
\ENDFOR
\FOR{each $c$ in 3-cells}
    \STATE $\text{PrepZ}(c)$
\ENDFOR
\FOR{each $\text{direction}$ in $\{+0,-0,+1,-1,+2,-2,+3,-3\}$}
    \FOR{each $c$ in 3-cells such that $c+\text{direction} = b$ in 2-cells}
            \STATE $\text{CNOT}(b, c)$
    \ENDFOR
\ENDFOR
\FOR{each $c$ in 3-cells}
    \STATE $\text{MeasZ}(c)$
\ENDFOR
\end{algorithmic}
\end{circuit}

\begin{circuit}
\caption{\label{compactse}Compact syndrome extraction circuit}
\begin{algorithmic}[1]
\FOR{each $a$ in 1-cells}
    \STATE $\text{PrepX}(a)$
\ENDFOR
\FOR{each $c$ in 3-cells}
    \STATE $\text{PrepZ}(c)$
\ENDFOR
\FOR{each $\text{direction}$ in $\{-3, -2, -1, -0, +0, +1, +2, +3\}$}
    \FOR{each $a$ in 1-cells such that $a+\text{direction} = b$ in 2-cells}
        \STATE $\text{CNOT}(a, b)$
    \ENDFOR
    \FOR{each $c$ in 3-cells such that $c+\text{direction} = b$ in 2-cells}
        \STATE $\text{CNOT}(b, c)$
    \ENDFOR
\ENDFOR
\FOR{each $a$ in 1-cells}
    \STATE $\text{MeasX}(a)$
\ENDFOR
\FOR{each $c$ in 3-cells}
    \STATE $\text{MeasZ}(c)$
\ENDFOR
\end{algorithmic}
\end{circuit}

\section{Lattices used in simulations}\label{app:simulation-code-parameters}

The performance simulations in \Cref{sec:memory-performance} used the HNFs in \Cref{tab:simulation-parameters}.

\begin{table}[hb]
\centering
\begin{tabular}{ccc}
\toprule
Hermite normal form & Determinant & Distance \\
\midrule
$\left[\begin{array}{rrrr}1& 0& 0& 1\\ 0& 1& 0& 1\\ 0& 0& 1& 1\\ 0& 0& 0& 3\end{array}\right]$ & 3 & 3 \\[8ex]
$\left[\begin{array}{rrrr}1& 0& 0& 5\\ 0& 1& 0& 6\\ 0& 0& 1& 7\\ 0& 0& 0& 9\end{array}\right]$ & 9 & 6 \\[8ex]
$\left[\begin{array}{rrrr}1& 0& 0& 4\\ 0& 1& 0& 6\\ 0& 0& 1& 7\\ 0& 0& 0& 9\end{array}\right]$ & 9 & 6 \\[8ex]
$\left[\begin{array}{rrrr}1& 1& 1& 1\\ 0& 2& 0& 2\\ 0& 0& 2& 2\\ 0& 0& 0& 4\end{array}\right]$ & 16* & 8 \\[8ex]
$\left[\begin{array}{rrrr}1& 0& 0& 3\\ 0& 1& 0& 5\\ 0& 0& 1& 7\\ 0& 0& 0& 16\end{array}\right]$ & 16 & 8 \\[8ex]
$\left[\begin{array}{rrrr}1& 0& 1& 6\\ 0& 1& 0& 11\\ 0& 0& 3& 9\\ 0& 0& 0& 15\end{array}\right]$ & 45 & 15 \\
\bottomrule
\end{tabular}
\caption{Hermite normal form, determinant, and distance for codes simulated in \Cref{sec:memory-performance}. The asterisk highlights the parameters for the Hadamard lattice.}
\label{tab:simulation-parameters}
\end{table}

\newpage

\bibliographystyle{plain}
\bibliography{references}

\section{Figures}

\end{document}